\documentclass[english,notitlepage,nofootinbib]{revtex4-1}
\usepackage[T1]{fontenc}
\usepackage[latin9]{inputenc}
\usepackage{color}
\usepackage{amsmath}
\usepackage{amssymb}
\usepackage{graphicx}
\usepackage{babel}
\usepackage{hyperref}
\begin{document}

\title{Trimaximal $\mu$-$\tau$ reflection symmetry}

\date{\today}

\author{Werner Rodejohann and Xun-Jie Xu }

\affiliation{\textcolor{black}{Max-Planck-Institut f\"ur Kernphysik, Postfach
103980, D-69029 Heidelberg, Germany}}
\begin{abstract}
\noindent The $\mu$-$\tau$ reflection symmetry $(\nu_{e},\thinspace\nu_{\mu},\thinspace\nu_{\tau})\rightarrow(\overline{\nu}_{e},\thinspace\overline{\nu}_{\tau},\thinspace\overline{\nu}_{\mu})$
and the TM1 mixing (a PMNS matrix with the first column fixed to the
TBM form) are both well compatible with experiments. If both
approaches are simultaneously assumed, all lepton mixing parameters
except for $\theta_{13}$ are predicted. In particular, one expects
maximal CP violation ($|\delta|=90^{\circ}$),
 maximal atmospheric mixing ($\theta_{23}=45^{\circ}$), a slightly
less-than-TBM solar mixing angle ($\theta_{12}\approx34^{\circ}$), as
well as values of $0$ or $\pi$ for the two Majorana phases.
We study the renormalization stability of
this highly predictive framework when neutrino mass is
described by an effective Weinberg operator and by
the type I seesaw mechanism, both in the Standard Model and with
supersymmetry.
\end{abstract}
\maketitle

\section{Introduction}
\noindent  The structure of neutrino mixing, the Pontecorvo\textendash Maki\textendash Nakagawa\textendash Sakata
(PMNS) matrix, is considered as an important clue for possible underlying
symmetries of the three generations of fermions in the Standard Model
(SM). Many discrete flavor symmetries have been proposed in trying
to understand the observed mixing -- see, e.g., the reviews \cite{Mohapatra:2006gs,Altarelli:2010gt,Ishimori:2010au,King:2013eh,Feruglio:2015jfa}.
In particular, it had long been speculated that the neutrino mixing
could be tribimaximal (TBM) \cite{Harrison:2002er,Harrison:2002kp,Xing:2002sw,Harrison:2002et,Harrison:2003aw},
which could originate from non-Abelian discrete symmetries such
as $A_{4}$ and $S_{4}$.  However, the TBM mixing  predicts zero
$\theta_{13}$ which has been excluded by reactor neutrino experiments
 \cite{Abe:2011fz,An:2012eh,Ahn:2012nd}.

It is well understood that $\theta_{13}=0$ in TBM is attributed to
$\mu$-$\tau$ symmetry \cite{Xing:2015fdg}, which is
defined as the invariance of
the neutrino mass terms under the interchange of $\nu_{\mu}$ and
$\nu_{\tau}$. Therefore in the light of non-zero $\theta_{13}$, breaking
the $\mu$-$\tau$ symmetry has been considered and extensively studied
in many references in the past. However, there is a variation
of the $\mu$-$\tau$ symmetry which does not require any breaking
and is still well compatible with experiments. It is called  $\mu$-$\tau$
reflection symmetry \cite{Harrison:2002et,Ma:2002ce,Babu:2002dz,Ma:2002ge,Grimus:2003yn},
which attaches the CP transformation to the interchange of $\nu_{\mu}$
and $\nu_{\tau}$,
\begin{equation}
\nu_{e}\rightarrow\overline{\nu}_{e},\thinspace\nu_{\mu}\rightarrow\overline{\nu}_{\tau},\thinspace\nu_{\tau}\rightarrow\overline{\nu}_{\mu}.\label{eq:mt-38}
\end{equation}
The $\mu$-$\tau$ reflection symmetry allows non-zero $\theta_{13}$
and predicts $\theta_{23}=45^{\circ}$ and $\delta=\pm90^{\circ}$.
Consequently
it has aroused a lot of interest recently \cite{Nishi:2013jqa, Ma:2013mga,
Fraser:2014yha,He:2015gba,Ma:2015gka,Li:2015jxa,DiIura:2015kfa,Mohapatra:2015gwa,
Zhou:2014sya,Joshipura:2015dsa,He:2015xha,
Zhao:2017yvw,Nishi:2016wki,Chen:2015siy,
Fukuyama:2017qxb}.
To generate TBM mixing the $\mu$-$\tau$
symmetry determines the third column of this mixing matrix and there
is another $\mathbb{Z}_{2}$ symmetry that is responsible for the first or second column
\cite{Lam:2008rs,Lam:2008sh,Lam:2011ag}. Those $\mathbb{Z}_{2}$ symmetries are assumed to be
``residual symmetries'', after the full flavor group is broken. They could be accidental or
subgroups of the full flavor group.
If the $\mu$-$\tau$ symmetry is replaced  with $\mu$-$\tau$ reflection symmetry,
then we get a variation of TBM with its first or second column
fixed and at the same time we will have non-zero $\theta_{13}$, $\theta_{23}=45^{\circ}$
and $\delta=\pm90^{\circ}$. We study the consequences of this assumption in this paper.
General deviations of the TBM mixing
with some part being fixed have been discussed in many references
\cite{Bjorken:2005rm,Xing:2006ms,He:2006qd,Albright:2008rp,Albright:2010ap,
Antusch:2011ic,He:2011gb,Varzielas:2012pa,Luhn:2013vna,Li:2013jya}
 and the case that the first/second column is fixed is usually referred
to as TM1/TM2 mixing, respectively \cite{Albright:2008rp}. In the TBM mixing, $\theta_{12}=\sin^{-1}\frac{1}{\sqrt{3}}\approx35.3^{\circ}$
is a little higher than the global best-fit value
$\theta_{12}^{{\rm exp}}=33.56_{-0.75}^{+0.77}$
\cite{Esteban:2016qun}, while in TM1 or TM2 it deviates from $35.3^{\circ}$
with a lower or a higher  value, respectively \cite{Albright:2008rp}:
\begin{equation}
\theta_{12}^{{\rm TM1}}=\cos^{-1}\left(\frac{\sqrt{2}}{\sqrt{3}\cos\theta_{13}}\right)\approx34.2^{\circ},\ \ \theta_{12}^{{\rm TM2}}=\sin^{-1}\left(\frac{1}{\sqrt{3}\cos\theta_{13}}\right)\approx35.8^{\circ}.\label{eq:mt-39}
\end{equation}
Since $\theta_{12}^{{\rm TM1}}$ is  well compatible with $\theta_{12}^{{\rm exp}}$
while $\theta_{12}^{{\rm TM2}}$ is disfavored at about $3\sigma$,
in this paper we will consider TM1 only.

When the TM1 symmetry\footnote{For simplicity, we will refer to the symmetry responsible for the
TM1 mixing as the TM1 symmetry in this paper.} and $\mu$-$\tau$ reflection symmetry are imposed on the neutrino
mass terms simultaneously, all the PMNS parameters except for $\theta_{13}$
are predicted (in addition to the predictions mentioned above, the
two Majorana phases are $0$ or $\pi$).
In the near future, this framework
can be tested not only by a precision measurement of $\theta_{12}$
and $\theta_{23}$,  but also by the confirmation of a maximal Dirac
CP phase $|\delta|=\pi/2$, for which hints have recently appeared
in T2K \cite{Abe:2013hdq,Abe:2017uxa}. Besides, its predictions on
Majorana phases could be verified in neutrinoless double beta decay
($0\nu\beta\beta$) experiments \cite{Rodejohann:2011mu}.


Note that both the TM1 symmetry and $\mu-\tau$ reflection symmetry
may be residual symmetries of a larger flavor symmetry broken
at a high energy scale. Since $\mu-\tau$ reflection
symmetry is essentially a generalized CP symmetry, looking for a horizontal
flavor symmetry that contains it as a subsymmetry is more interesting
and also more complicated. This is an active subject of on-going
research and some non-Abelian discrete groups in semidirect product
form, such as $A_{4}\rtimes\mathbb{Z}_{2}^{CP}$, $S_{4}\rtimes\mathbb{Z}_{2}^{CP}$,
$\Delta(6n^{2})\rtimes\mathbb{Z}_{2}^{CP}$ can be the origin of the
mixing scheme that we study here \cite{Feruglio:2012cw,Feruglio:2013hia,Ding:2014ora,Li:2016nap}.

It is most likely that the predictions of TM1 and $\mu-\tau$ reflection
symmetries are exact only at the scale where the horizontal flavor
symmetry breaks into these residual symmetries.
When going to lower energy scales
these predictions will unavoidably receive corrections
from renormalization group (RG) running \cite{Ohlsson:2013xva}. Therefore
in this paper, we will also study the RG corrections on the predictions
from the joined TM1 and $\mu$-$\tau$ reflection symmetry. We consider the case in which
neutrino mass is described by the effective Weinberg operator, as well as by the
most popular realization of this operator, the type I
seesaw \cite{Minkowski:1977sc,yanagida1979proceedings,glashow1979future,mohapatra1980neutrino}.
Both the SM and the (Minimal Supersymmetric Standard Model) MSSM are assumed. \\

The remainder of the paper is organized as follows. In Sec.\ \ref{sec:basic},
we introduce the TM1 symmetry and the $\mu$-$\tau$ reflection symmetry, and study the
phenomenology if both are simultaneously present.
Then we study the RG running effects on the PMNS parameters in the cases we
mentioned above, presented in Sec.\ \ref{sec:RG}. Finally we summarize
our result and conclude in Sec.\ \ref{sec:Conclusion}.

\section{Trimaximal $\mu$-$\tau$ reflection symmetry\label{sec:basic}}
\noindent
The TM1 mixing and its symmetry as well as model-building aspects have been
studied in many references (see e.g.\ \cite{Albright:2008rp,Varzielas:2012pa,Luhn:2013vna,Li:2013jya,Rodejohann:2012cf,Shimizu:2014ria,Ballett:2016yod}).
 In the following we denote the TM1 symmetry as  $\mathbb{Z}_{2}^{{\rm TM1}}$.
The $\mu$-$\tau$ reflection symmetry was originally proposed in
Refs.\ \cite{Harrison:2002et,Ma:2002ce,Babu:2002dz,Ma:2002ge,Grimus:2003yn}
and later extensively studied in, e.g.,
\cite{Nishi:2013jqa, Ma:2013mga,
Fraser:2014yha,He:2015gba,Ma:2015gka,DiIura:2015kfa,Mohapatra:2015gwa,
Zhou:2014sya,Joshipura:2015dsa,He:2015xha,
Zhao:2017yvw,Nishi:2016wki,Chen:2015siy,
Fukuyama:2017qxb}.
It can be regarded as a generalized CP symmetry \cite{Neufeld:1987wa,Grimus:1995zi}
so we use $\mathbb{Z}_{2}^{\rm CP}$ to denote it\footnote{We prefer the symbol $\mathbb{Z}_{2}^{\rm CP}$ to $\mathbb{Z}_{2}^{\mu\tau}$
for the $\mu$-$\tau$ reflection symmetry because the latter is widely
used for the $\mu$-$\tau$ symmetry without CP transformation.}.
Although both symmetries as well as their phenomenology have been
extensive studied in the literature, their combination which provides
a very effective description of the neutrino mixing data with only
one free parameter, has attracted much less attention. Therefore in
this section, we will discuss the theoretical and phenomenological
aspects of this combination.

The explicit transformations of $\mathbb{Z}_{2}^{{\rm TM1}}$ and
$\mathbb{Z}_{2}^{\rm CP}$ in the flavor basis are given as
\begin{equation}
{\rm \mathbb{Z}_{2}^{TM1}:\ }\nu\rightarrow R^{{\rm TM1}}\nu,\label{eq:mt}
\end{equation}
\begin{equation}
{\rm \mathbb{Z}_{2}^{\rm CP}:\ }\nu\rightarrow R^{\mu\tau}\overline{\nu},\label{eq:mt-2}
\end{equation}
where $\nu\equiv(\nu_{e},\thinspace\nu_{\mu},\thinspace\nu_{\tau})^{T}$
and the two matrices $R^{{\rm TM1}}$ and $R^{\mu\tau}$ are
\begin{equation}
R^{{\rm TM1}}\equiv-\frac{1}{3}\left(\begin{array}{ccc}
1 & 2 & 2\\
2 & -2 & 1\\
2 & 1 & -2
\end{array}\right),\label{eq:mt-1}
\end{equation}
\begin{equation}
R^{\mu\tau}\equiv\left(\begin{array}{ccc}
1 & 0 & 0\\
0 & 0 & 1\\
0 & 1 & 0
\end{array}\right).\label{eq:mt-3}
\end{equation}
The matrix $R^{{\rm TM1}}$  has been derived in, e.g., Ref. \cite{Lam:2008sh}
while the form of $R^{\mu\tau}$ is obvious according to  the meaning
of interchanging the $\mu$ and $\tau$ flavor. Since the $\mu$-$\tau$
reflection symmetry is essentially a generalized CP symmetry, it
is necessary to check the consistency condition of flavor symmetry
and CP symmetry \cite{Holthausen:2012dk}:
\begin{equation}
R^{{\rm TM1}}R^{\mu\tau}=R^{\mu\tau}(R^{{\rm TM1}})^{*}.\label{eq:mt-4}
\end{equation}

The neutrino mass terms
\begin{equation}
{\cal L}\supset-\nu_{\alpha}M_{\alpha\beta}^{\nu}\nu_{\beta}+{\rm h.c.,}\label{eq:mt-5}
\end{equation}
should be invariant under the transformations in Eqs.~(\ref{eq:mt})
and (\ref{eq:mt-2}). Therefore, the mass matrix $M^{\nu}$ should
satisfy
\begin{equation}
(R^{{\rm TM1}})^{T}M^{\nu}R^{{\rm TM1}}=M^{\nu},\label{eq:mt-6}
\end{equation}
\begin{equation}
(R^{\mu\tau})^{T}M^{\nu}R^{\mu\tau}=(M^{\nu})^{*}.\label{eq:mt-7}
\end{equation}
The above two equations
can be broken down into equations in terms of the entries of
$M^{\nu}$, so one can obtain explicit constraints on those:
\begin{eqnarray}
M_{11}^{\nu},\thinspace M_{23}^{\nu} & = & {\rm real},\label{eq:mt-8}\\
M_{12}^{\nu} & = & (M_{13}^{\nu})^{*},\label{eq:mt-9}\\
M_{22}^{\nu} & = & (M_{33}^{\nu})^{*},\label{eq:mt-10}\\
{\rm Im}(M_{22}^{\nu}) & = & 2{\rm Im}(M_{23}^{\nu}),\label{eq:mt-11}\\
M_{11}^{\nu} & = & \sum_{i}{\rm Re}(M_{i2}^{\nu}).\label{eq:mt-12}
\end{eqnarray}
The above equations are equivalent to Eqs.~(\ref{eq:mt-6}) and (\ref{eq:mt-7}),
which means they are sufficient and necessary conditions for Eq.~(\ref{eq:mt-5})
being invariant under the transformations. With the above constraints,
$M^{\nu}$ can be parametrized by four real parameters $r$, $x_{1,2}$
and $y$:
\begin{equation}
M^{\nu}=\left(\begin{array}{ccc}
r+x_{1}+x_{2} & x_{1} & x_{1}\\
x_{1} & x_{2} & r\\
x_{1} & r & x_{2}
\end{array}\right)+iy\left(\begin{array}{ccc}
0 & 1 & -1\\
1 & -2 & 0\\
-1 & 0 & 2
\end{array}\right).\label{eq:mt-13}
\end{equation}
As one can check, Eq.~(\ref{eq:mt-13}) is the most general mass matrix
that satisfied Eqs.~(\ref{eq:mt-6}) and (\ref{eq:mt-7}). The
mass matrix contains only four real parameters; those are the three neutrinos
masses and one degree of freedom for the PMNS matrix. As
we will show later, this degree of freedom is just $\theta_{13}$.
Therefore, the mass matrix with the form in Eq.~(\ref{eq:mt-13})
is highly predictive. It predicts all the parameters except for $\theta_{13}$
in the PMNS matrix, including two mixing angles $(\theta_{12},\thinspace\theta_{23})$,
one Dirac phase $\delta$ and two Majorana phases $(\alpha_{21},\thinspace\alpha_{31})$.

The mass matrix is diagonalized by
\begin{equation}
(U^{\nu})^{T}M^{\nu}U^{\nu}={\rm diag}(m'_{1},\thinspace m'_{2},\thinspace m'_{3}).\label{eq:mt-14}
\end{equation}
For $M^{\nu}$ in Eq.~(\ref{eq:mt-13}), due to the residual symmetries,
$U^{\nu}$ can be analytically solved:
\begin{equation}
U^{\nu}=\frac{1}{\sqrt{6}}\left(\begin{array}{ccc}
2 & \sqrt{2}c & \sqrt{2}s\\
1 & -\sqrt{2}c-i\sqrt{3}s & i\sqrt{3}c-\sqrt{2}s\\
1 & -\sqrt{2}c+i\sqrt{3}s & -i\sqrt{3}c-\sqrt{2}s
\end{array}\right),\label{eq:mt-15}
\end{equation}
where $(s,\thinspace c)=(\sin\theta,\thinspace\cos\theta)$ are given
by
\begin{equation}
s=\sqrt{\frac{\Delta+x_{1}-2x_{2}}{2\Delta}},\ c={\rm sign}(y)\sqrt{1-\frac{x_{1}-2x_{2}+\Delta}{2\Delta}},\label{eq:mt-16}
\end{equation}
and
\begin{equation}
(m'_{1},\thinspace m'_{2},\thinspace m'_{3})=\left(r+2x_{1}+x_{2},\thinspace r+\frac{\Delta}{2}-\frac{x_{1}}{2},\thinspace r-\frac{\Delta}{2}-\frac{x_{1}}{2}\right),\label{eq:mt-17}
\end{equation}
\begin{equation}
\Delta\equiv\sqrt{24y^{2}+\left(x_{1}-2x_{2}\right){}^{2}}.\label{eq:mt-18}
\end{equation}
Here ${\rm sign}(y)$ implies that we have taken $c=\sqrt{1-s^{2}}$
for positive $y$ and $c=-\sqrt{1-s^{2}}$ for negative $y$. Note
that $(m'_{1},\thinspace m'_{2},\thinspace m'_{3})$ computed from
Eq.~(\ref{eq:mt-17}) are not necessarily positive (but always real),
so they may be different from the neutrino masses by some minus signs.

Comparing the above result to the standard parametrization of the
PMNS matrix
\begin{equation}
U_{{\rm PMNS}}=U\thinspace{\rm diag}(1,\thinspace e^{i\alpha_{21}/2},\thinspace e^{i\alpha_{31}/2}),\label{eq:mt-19}
\end{equation}
\begin{equation}
U=\left(\begin{array}{ccc}
c_{12}c_{13} & c_{13}s_{12} & e^{-i\delta}s_{13}\\
-c_{23}s_{12}-e^{i\delta}c_{12}s_{13}s_{23} & c_{12}c_{23}-e^{i\delta}s_{12}s_{13}s_{23} & c_{13}s_{23}\\
-e^{i\delta}c_{12}c_{23}s_{13}+s_{12}s_{23} & -e^{i\delta}c_{23}s_{12}s_{13}-c_{12}s_{23} & c_{13}c_{23}
\end{array}\right),\label{eq:mt-20}
\end{equation}
we can extract the predictions on all the PMNS parameters. It turns
out that the predictions differ for positive and negative $y$. Next we
will discuss both cases:
\begin{itemize}
\item Positive $y$ ($y>0$)\newline
If $y>0$, then $c=\sqrt{1-s^{2}}$ is positive.  We extract some
phases from $U^{\nu}$ so that
\begin{equation}
{\rm diag}(1,\thinspace-e^{i\beta},\thinspace e^{-i\beta})U^{\nu}=\left(\begin{array}{ccc}
\sqrt{\frac{2}{3}} & \frac{\sqrt{1-s^{2}}}{\sqrt{3}} & \frac{is}{\sqrt{3}}\\
\frac{2i\sqrt{3}s-3\sqrt{2-2s^{2}}}{6\sqrt{3-s^{2}}} & \frac{6+is\sqrt{6-6s^{2}}}{6\sqrt{3-s^{2}}} & \frac{\sqrt{3-s^{2}}}{\sqrt{6}}\\
\frac{2i\sqrt{3}s+3\sqrt{2-2s^{2}}}{6\sqrt{3-s^{2}}} & \frac{-6+is\sqrt{6-6s^{2}}}{6\sqrt{3-s^{2}}} & \frac{\sqrt{3-s^{2}}}{\sqrt{6}}
\end{array}\right){\rm diag}(1,\thinspace1,\thinspace-i),\label{eq:mt-21}
\end{equation}
has the same phase convention as the standard parametrization, which
requires
\begin{equation}
\beta=\arg(\sqrt{3}c-i\sqrt{2}s).\label{eq:mt-22}
\end{equation}
Comparing Eq.~(\ref{eq:mt-21}) to Eqs.~(\ref{eq:mt-20}) and (\ref{eq:mt-19}),
we get
\begin{equation}
\theta_{23}=45^{\circ},\thinspace\delta=-90^{\circ},\thinspace c_{12}=\sqrt{\frac{2}{3}}\frac{1}{c_{13}}.\label{eq:mt-23}
\end{equation}
 If $m'_{1,2,3}\geq0$, then the Majorana phases should be $(1,\thinspace e^{i\alpha_{21}/2},\thinspace e^{i\alpha_{31}/2})=(1,\thinspace1,\thinspace-i)$.
However, $m'_{1,2,3}$ could be negative, which can be converted to
positive by further adding some phases to the right-hand side of Eq.~(\ref{eq:mt-21}).
Therefore the actual Majorana phases depend on the signs of $m'_{1,2,3}$:
\begin{equation}
(1,\thinspace e^{i\alpha_{21}/2},\thinspace e^{i\alpha_{31}/2})=\left(1,\thinspace\sqrt{{\rm sign}(m'_{2}/m'_{1})},\thinspace-i\sqrt{{\rm sign}(m'_{3}/m'_{1})}\right).\label{eq:mt-24}
\end{equation}

\item Negative $y$ ($y<0$)\newline
If $y<0$, then $c=-\sqrt{1-s^{2}}$ is negative so we need to remove
the minus sign of the 12-entry of (\ref{eq:mt-15}). Therefore Eqs.~(\ref{eq:mt-21})
and (\ref{eq:mt-22}) are modified to
\begin{equation}
{\rm diag}(1,\thinspace-e^{i\beta},\thinspace e^{-i\beta})U^{\nu}=\left(\begin{array}{ccc}
\sqrt{\frac{2}{3}} & \frac{\sqrt{1-s^{2}}}{\sqrt{3}} & -\frac{is}{\sqrt{3}}\\
\frac{-2i\sqrt{3}s-3\sqrt{2-2s^{2}}}{6\sqrt{3-s^{2}}} & \frac{6-is\sqrt{6-6s^{2}}}{6\sqrt{3-s^{2}}} & \frac{\sqrt{3-s^{2}}}{\sqrt{6}}\\
\frac{-2i\sqrt{3}s+3\sqrt{2-2s^{2}}}{6\sqrt{3-s^{2}}} & \frac{-6-is\sqrt{6-6s^{2}}}{6\sqrt{3-s^{2}}} & \frac{\sqrt{3-s^{2}}}{\sqrt{6}}
\end{array}\right){\rm diag}(1,\thinspace-1,\thinspace i),\label{eq:mt-25}
\end{equation}
where now
\begin{equation}
\beta=\arg(-\sqrt{3}c+i\sqrt{2}s).\label{eq:mt-26}
\end{equation}
In this case, comparing with the standard parametrization of the PMNS matrix we have
\begin{equation}
\theta_{23}=45^{\circ},\thinspace\delta=90^{\circ},\thinspace c_{12}=\sqrt{\frac{2}{3}}\frac{1}{c_{13}},\label{eq:mt-27}
\end{equation}
and the Majorana phases are
\[
(1,\thinspace e^{i\alpha_{21}/2},\thinspace e^{i\alpha_{31}/2})=\left(1,\thinspace\sqrt{{\rm sign}(m'_{2}/m'_{1})},\thinspace i\sqrt{{\rm sign}(m'_{3}/m'_{1})}\right).
\]
\end{itemize}

\begin{figure}
\centering

\includegraphics[width=10cm]{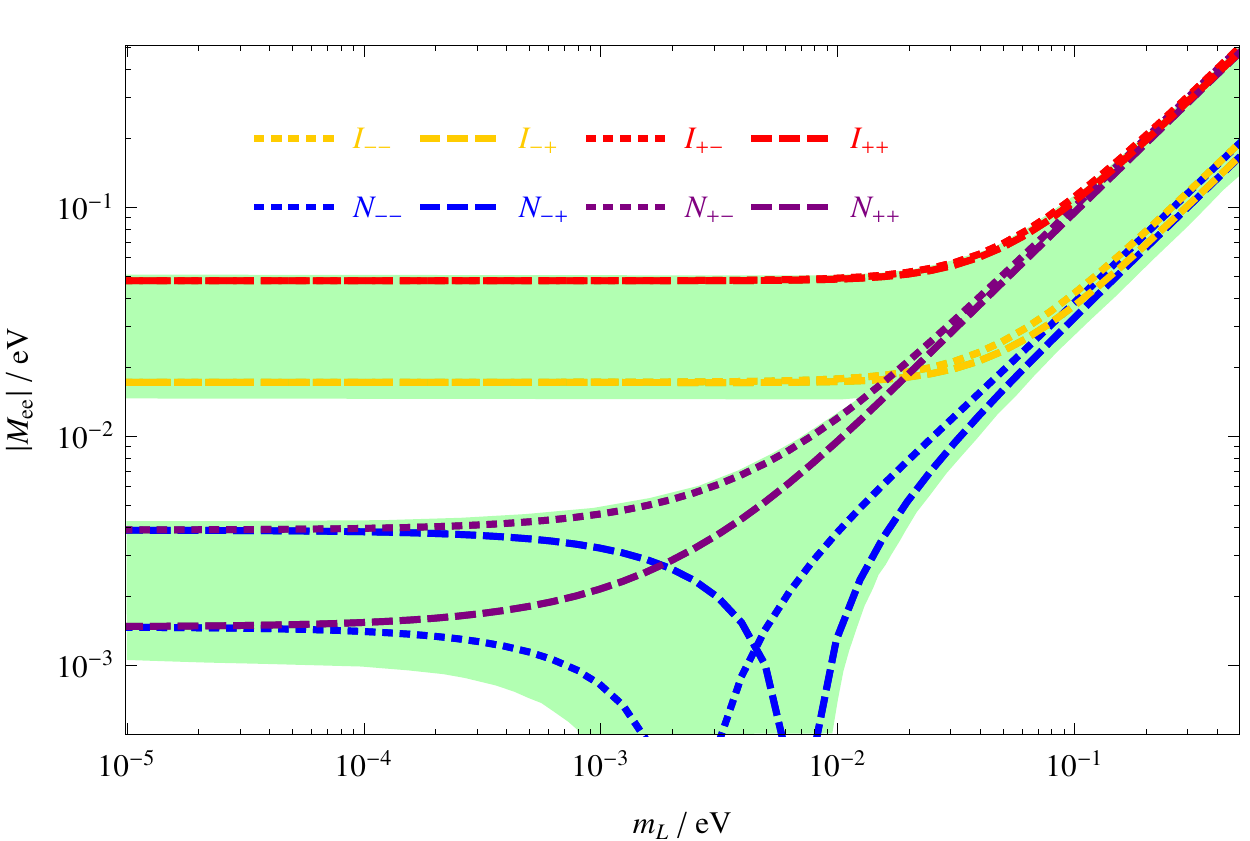}\,\caption{\label{fig:mee}Prediction
  on the  effective mass
$|M_{ee}|$, according to Eqs.~(\ref{eq:mt-28}) and (\ref{eq:mt-29}).
In the notation $N_{\pm\pm}$ ($I_{\pm\pm}$), $N/I$ standards for
normal/inverted mass ordering respectively and the subscripts are
the signs of two Majorana phases $e^{i\alpha_{21}}$ and $e^{i\alpha_{31}}$
($\alpha_{21}$, $\alpha_{31}$ are always 0 or $\pi$ in this model).
The light green region is the bound from the global fit, taken from
\cite{Olive:2016xmw}.}
\end{figure}

As a summary, we have
\begin{equation}
\theta_{23}=45^{\circ},\ \theta_{12}=\cos^{-1}(\sqrt{\frac{2}{3}}\frac{1}{c_{13}})\approx34.2^{\circ},\thinspace\label{eq:mt-28}
\end{equation}
if the experimental value $\theta_{13}\approx9^{\circ}$ is taken
as an input, and
\begin{equation}
\delta=\pm90^{\circ},\ \alpha_{21}=\frac{\pi}{2}\pm\frac{\pi}{2},\ \alpha_{31}=\frac{\pi}{2}\pm\frac{\pi}{2},\label{eq:mt-29}
\end{equation}
where the positive/negative signs depending on the signs of $y$
and $(m'_{1},\thinspace m'_{2},\thinspace m'_{3})$ computed from
Eq.~(\ref{eq:mt-17}). Here the Majorana phases are predicted to be
either $0$ or $\pi$. There have been many studies
\cite{Rodejohann:2000ne,Ge:2016tfx,Bilenky:2001rz,
Pascoli:2002qm,Minakata:2014jba,Simkovic:2012hq,
Joniec:2004mx,
Rodejohann:2002ng,Benato:2015via}
on the option to measure the Majorana phases with upcoming
neutrinoless double beta decay ($0\nu\beta\beta$) experiments.  It was demonstrated in particular
that expected nuclear and experimental uncertainties allow in principle
to measure the phases, or at least contrain them non-trivially.
The actual physical observable
for $0\nu\beta\beta$ is the effective mass $|M_{ee}|$, which
has significant dependence on the Majorana phases. For the inverted
mass ordering, $|M_{ee}|$ is always nonzero, which necessarily leads
to $0\nu\beta\beta$ at some level. For the normal mass ordering, it is well
known that $|M_{ee}|$ can be zero for very small neutrino mass;
however, $|M_{ee}|=0$ does not mean
that $0\nu\beta\beta$ experiments tell us
nothing about the Majorana phases.
 As it has been noticed in Refs.\ \cite{Ge:2016tfx,Xing:2003jf}, this
case still gives some constraints on the Majorana phases. In the scenario
of this work, the relation between $|M_{ee}|$ and the Majorana phases
is more explicit because all the neutrino parameters except for the
lightest neutrino mass $m_{L}$ have been determined by symmetries
or by experiments, enabling us to compute $|M_{ee}|$ explicitly, as
shown in Fig.\ \ref{fig:mee}. Note that in this scenario, $|M_{ee}|<10^{-3}$
eV is possible only if $\alpha_{21}=\pi$. So if the future experiments
push the upper bound of $|M_{ee}|$ down to $10^{-3}$ eV and still
do not observe $0\nu\beta\beta$ decay, then we can draw the conclusion
that $\alpha_{21}=\pi$.



We can confront the predictions of the mixing scheme with current data \cite{Esteban:2016qun}. First we study
the predictions of TM1 mixing, namely the first column of the PMNS matrix being
$(\sqrt \frac 23$, $\sqrt \frac 16, \sqrt \frac 16)^T$.
The $\chi^2$-function is defined as
\begin{equation}
\chi^2 = \sum \frac{(x_i - x_i^0)^2}{\sigma_i^2},
\end{equation}
where $x_i^0$ represents the data of the $i$-th experimental
observable, $\sigma_i$ the corresponding $1\sigma$ absolute error, and
$x_i$ the prediction of the model. For the normal ordering, TM1 has a $\chi^2$-minimum of
$1.14  \, (=0.063+0.000+1.058+0.0223)$ at the values $\theta_{13} = 8.5^\circ$ and $\theta_{23} = 41.6^\circ$.
The numbers in brackets denote the contributions of $\theta_{13}, \theta_{23}, \theta_{12}$ and $\delta$ to
the total value. In case of an inverted ordering, the  $\chi^2$-minimum is
$1.20  \, (=0.006+0.000+1.056+0.143)$ at the values $\theta_{13} = 8.5^\circ$ and $\theta_{23} = 50.0^\circ$.
Note that TM1 has two free parameters.
Combining TM1 with $\mu$-$\tau$ reflection symmetry, which in total has only one free parameter,
gives for the
normal ordering a $\chi^2$-minimum of
$3.88  \, (=0.063+2.730+1.058+0.0308)$ at the value  $\theta_{13} = 8.5^\circ$. In the inverted ordering, the
 $\chi^2$-minimum is $5.76 \, (=0.006+4.672+1.056+0.0234)$ at the value $\theta_{13} = 8.5^\circ$.

\section{RG corrections\label{sec:RG}}
The residual symmetries we discussed in the previous section may appear
at a very high energy scale, which we refer to as the flavor symmetry
scale. Due to radiative corrections, the predictions at the flavor
symmetry scale may be modified at the low energy scale, at which they
are confronted with experimental measurements.  If there is no new
physics between the two scales, the corrections can be computed without
many unknown parameters involved. However, it is also possible that
some new physics appear in the middle so that the RG corrections would
depend on more unknown parameters. For example, in the type I seesaw
mechanism, the masses of right-handed neutrinos could be below the
flavor symmetry scale; in this case the RG corrections would also depend on
 the masses of right-handed neutrinos.

\subsection{RG running based on the Weinberg operator \label{sec:RG-Weinb}}

To avoid the dependence on too many parameters, we will first focus
on the case that all other new physics scales are above the flavor
symmetry scale. In this case, the calculation will be based on the
RGE of the SM extended by the Weinberg operator,
\begin{equation}
{\cal L}\supset\frac{1}{4}\kappa_{\alpha\beta}(\widetilde{H}^{\dagger}L_{\alpha})(\widetilde{H}^{\dagger}L_{\beta})+{\rm h.c.},\label{eq:mt-30}
\end{equation}
where $L$ is the lepton doublet and $H$ the Higgs doublet. After
electroweak symmetry breaking $\langle \tilde{H}\rangle=(v/\sqrt{2},0)^{T}$,
the neutrino mass matrix is given by
\begin{equation}
M_{\alpha\beta}^{\nu}=-\frac{v^{2}}{4}\kappa_{\alpha\beta}.\label{eq:mt-31}
\end{equation}
Constrained by the residual symmetries, $M^{\nu}$ depends on four
parameters $(r,\thinspace x_{1},\thinspace x_{2},\thinspace y)$ in
Eq.~(\ref{eq:mt-13}). Those parameters are actually highly constrained
by neutrino oscillation measurements on the two mass-squared differences
\begin{equation}
\delta m^{2}\equiv m_{2}^{2}-m_{1}^{2},\ \Delta m^{2}\equiv m_{3}^{2}-\frac{m_{1}^{2}+m_{2}^{2}}{2},\label{eq:mt-32}
\end{equation}
and $\sin\theta_{13}$. In this section we will fix them at the
best-fit
values \cite{Esteban:2016qun,Capozzi:2016rtj} as the result of
our calculation varies very little within experimental uncertainties.
If the lightest neutrino mass $m_{L}$ is also known, then $(r,\thinspace x_{1},\thinspace x_{2},\thinspace y)$
can be determined by $(\theta_{13},\thinspace\delta m^{2},\thinspace\Delta m^{2},\thinspace m_{L})$.
In Sec.\ \ref{sec:basic} we have demonstrated how to compute $(\theta_{13},\thinspace\delta m^{2},\thinspace\Delta m^{2},\thinspace m_{L})$
for given values of $(r,\thinspace x_{1},\thinspace x_{2},\thinspace y)$.
Determining $(r,\thinspace x_{1},\thinspace x_{2},\thinspace y)$
from experimental values of $(\theta_{13},\thinspace\delta
m^{2},\thinspace\Delta m^{2},\thinspace m_{L})$
is then of course also possible.

However there are some positive/negative signs one needs to choose
in determining $(r,\thinspace x_{1},\thinspace x_{2},\thinspace y)$.
The first one is the sign of $\Delta m^{2}$, known as the neutrino
mass ordering. Both   the normal  (NO, $\Delta m^{2}>0$)
and the inverted ordering (IO, $\Delta m^{2}<0$ ) should be taken
into consideration.   The next one is the sign of the Dirac phase
$\delta$. The $\mu$-$\tau$ reflection symmetry only predicts $|\delta|=90^{\circ}$
but both $+90^{\circ}$ and $-90^{\circ}$ are possible. Besides,
as summarized in Eq.~(\ref{eq:mt-29}), the two Majorana phases take
values of $\frac{\pi}{2}\pm\frac{\pi}{2}$, where we have to choose
between the positive/negative signs.

\begin{figure}
\centering

\includegraphics[width=7.5cm]{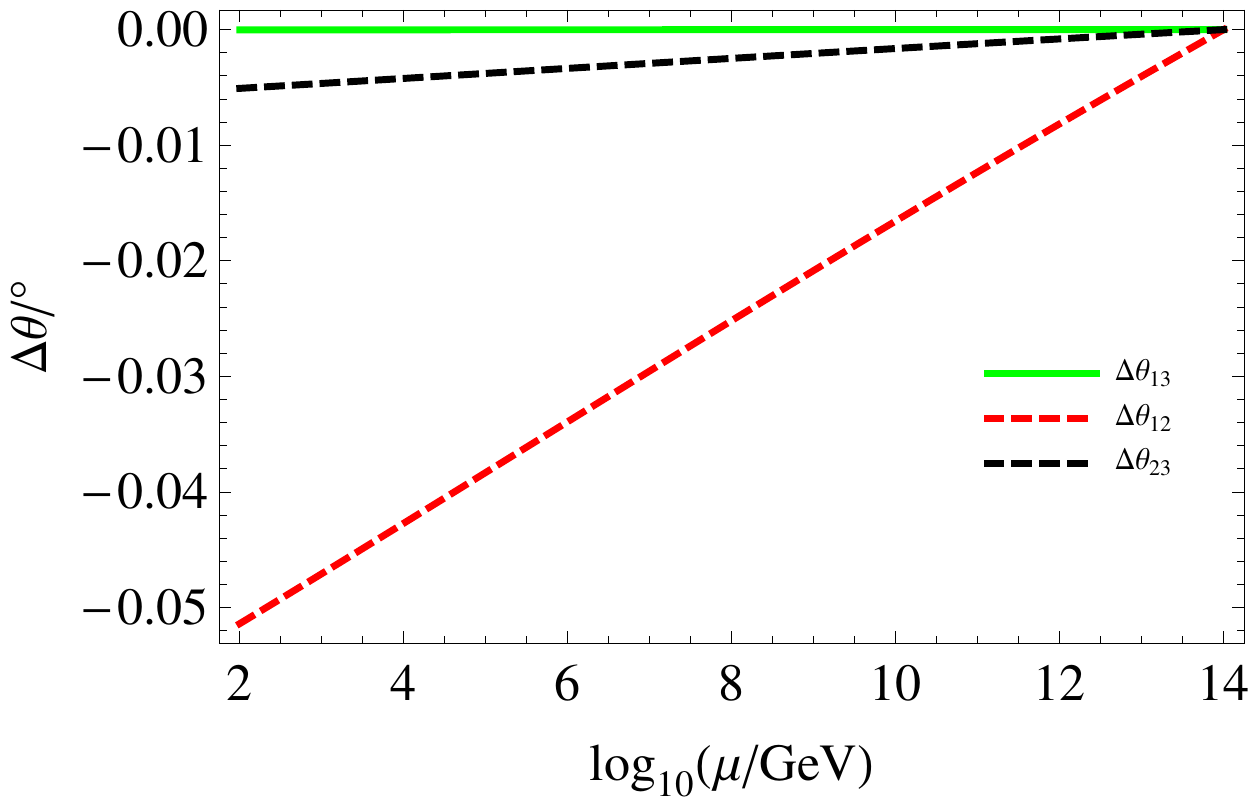}\,\includegraphics[width=7.5cm]{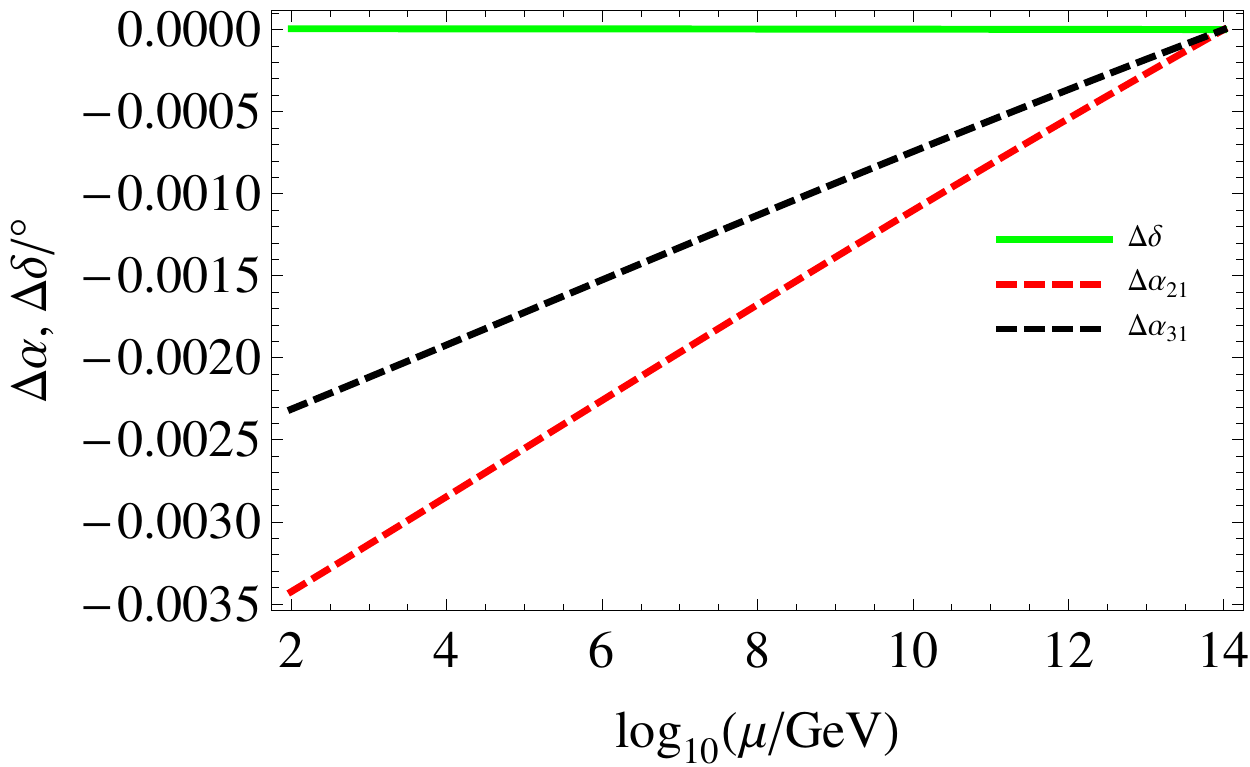}\caption{\label{fig:running}RG running of the mixing angles (left panel) and
the Dirac/Majorana phases (right panel) in the SM for the normal hierarchy and $m_{L}=0.05$
eV.}
\end{figure}

\begin{figure}
\centering

\begin{minipage}{0.42\textwidth}
\flushright
\includegraphics[height=4cm]{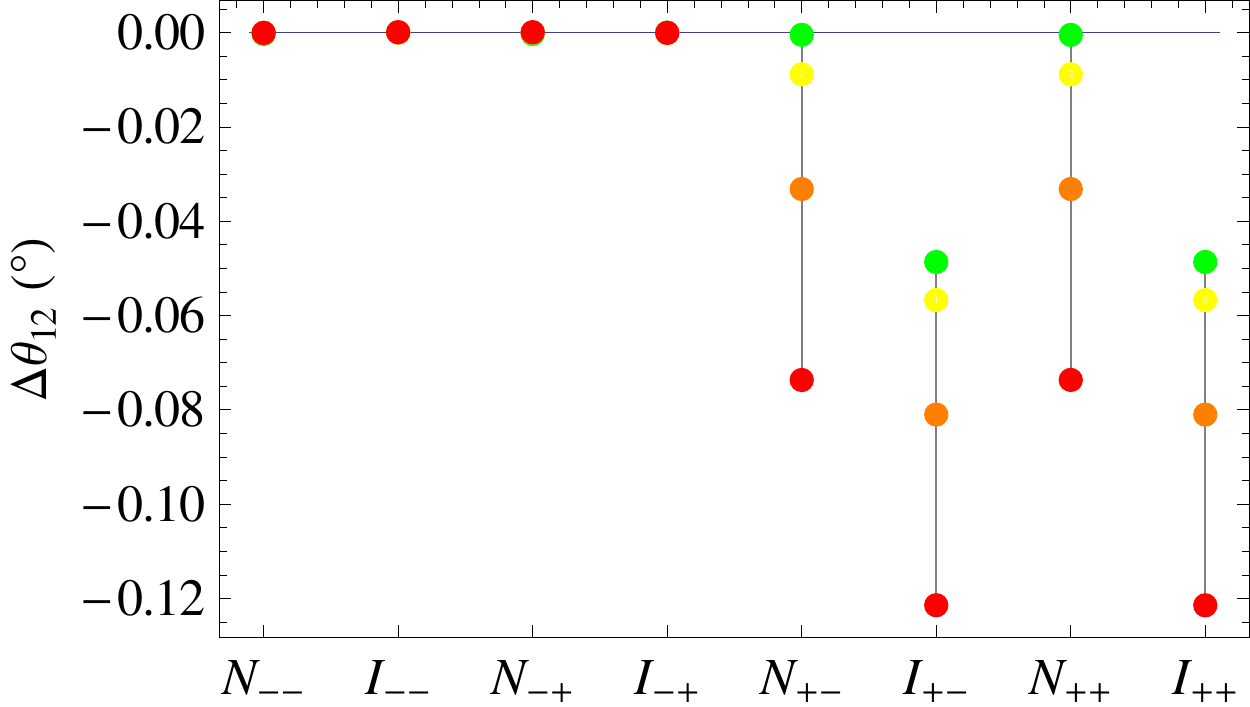}\\
\includegraphics[height=4cm]{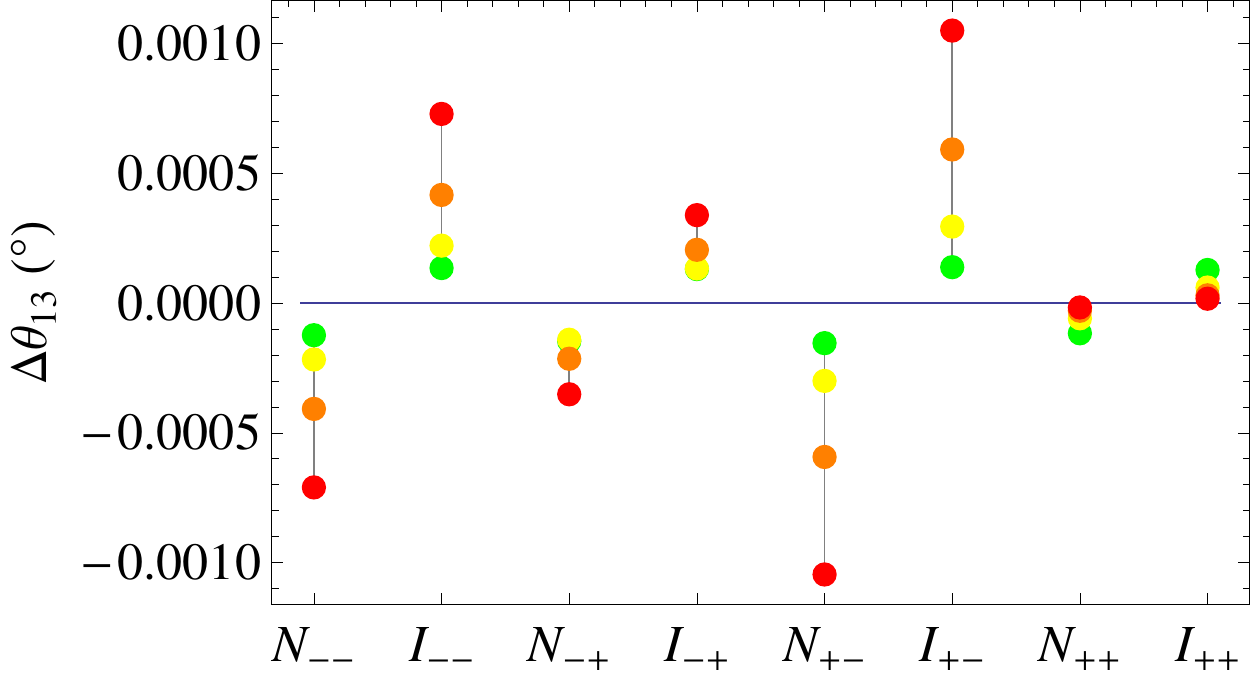}\\
\includegraphics[height=4cm]{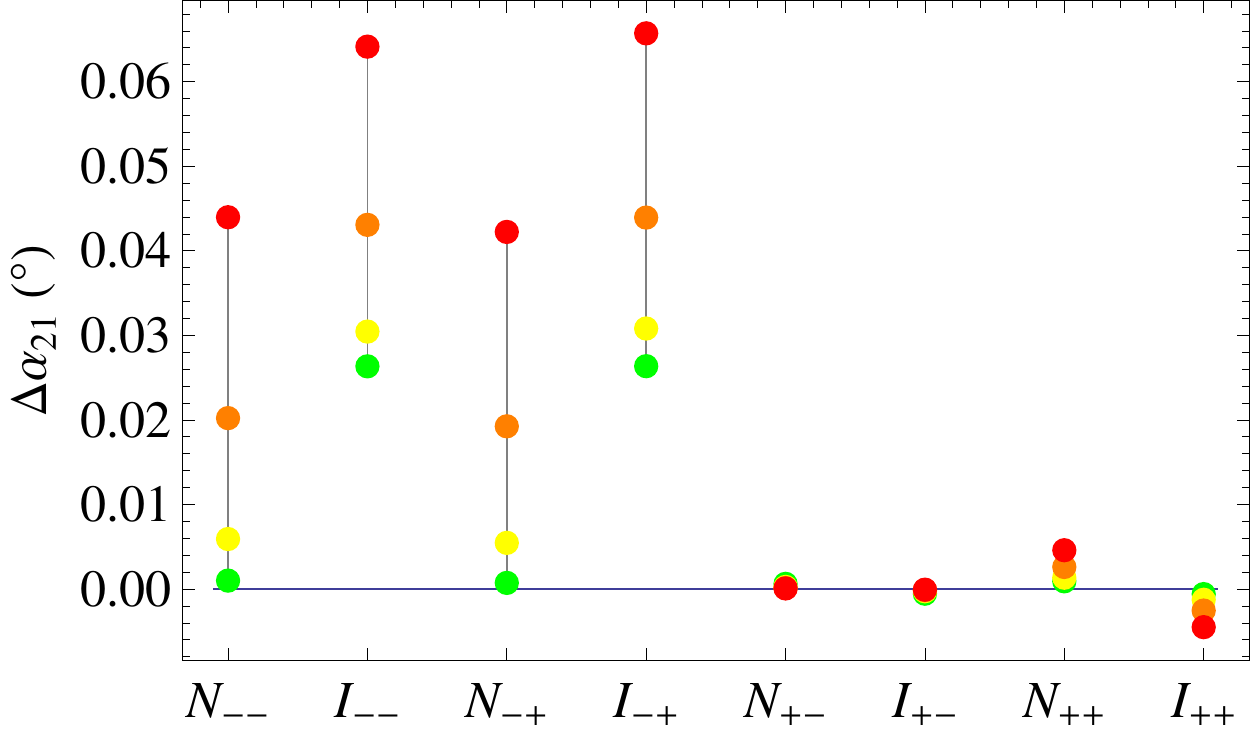}
\end{minipage}
\begin{minipage}{0.42\textwidth}
\flushright
\includegraphics[height=4cm]{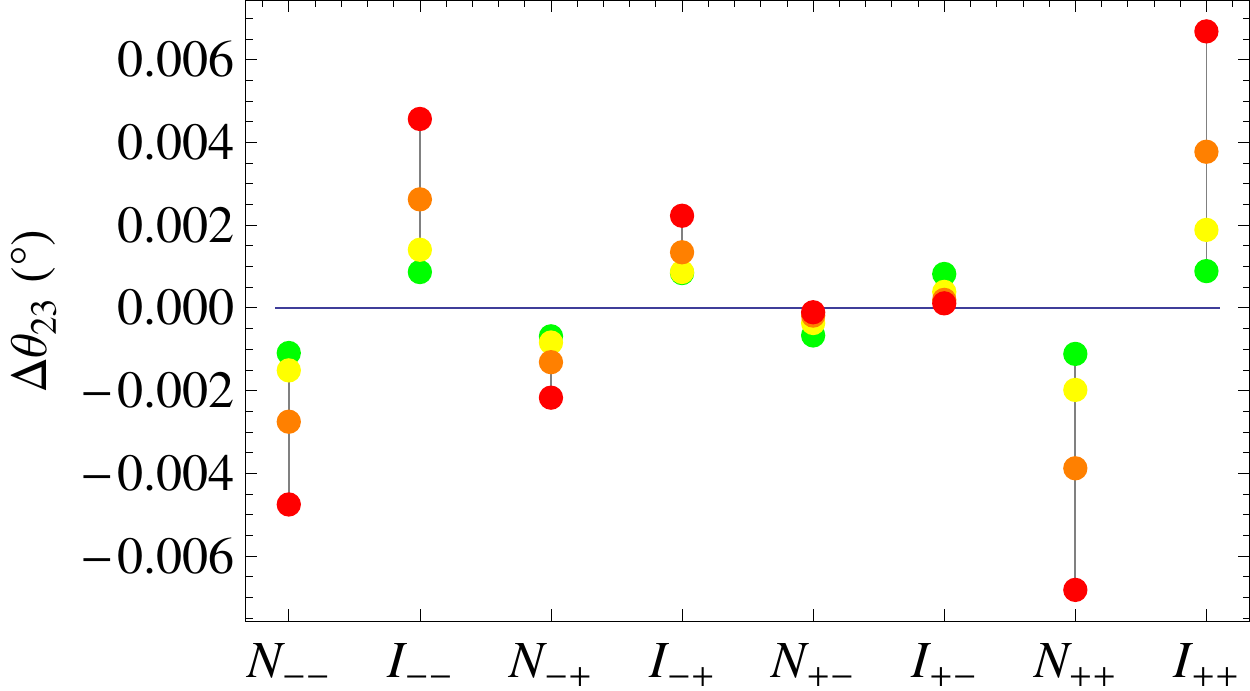}\\
\includegraphics[height=4cm]{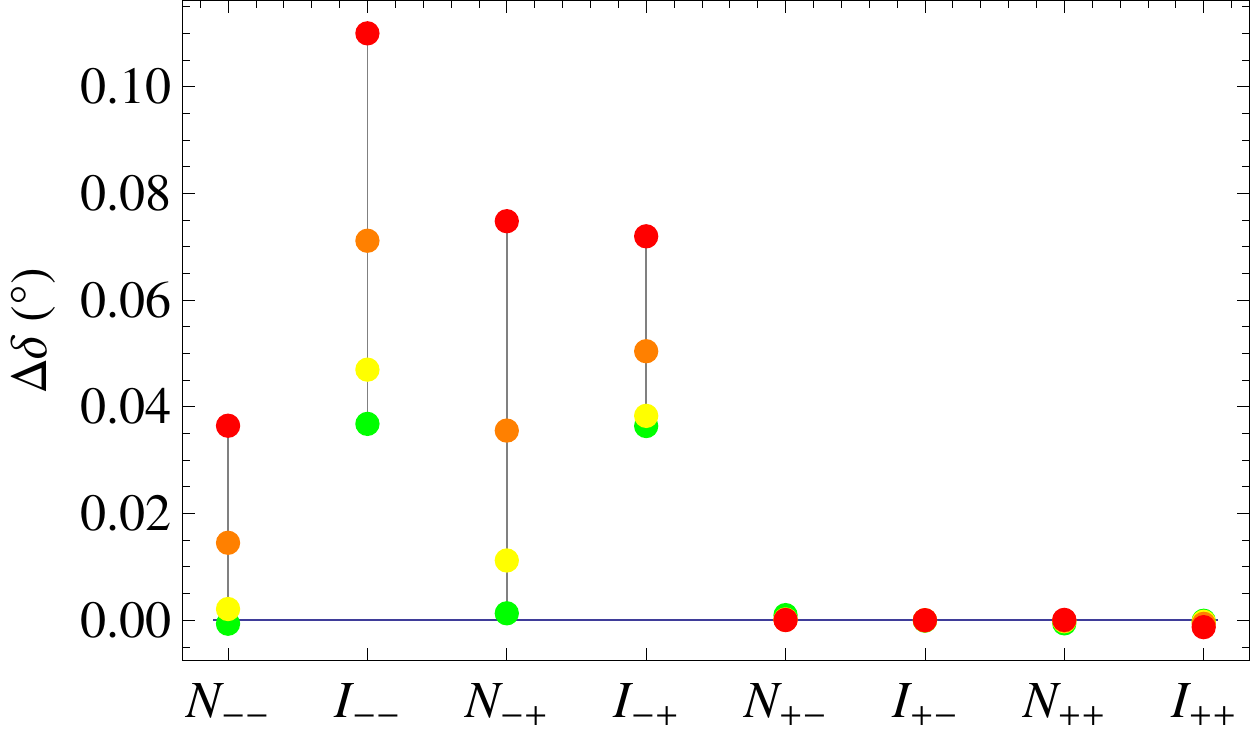}\\
\includegraphics[height=4cm]{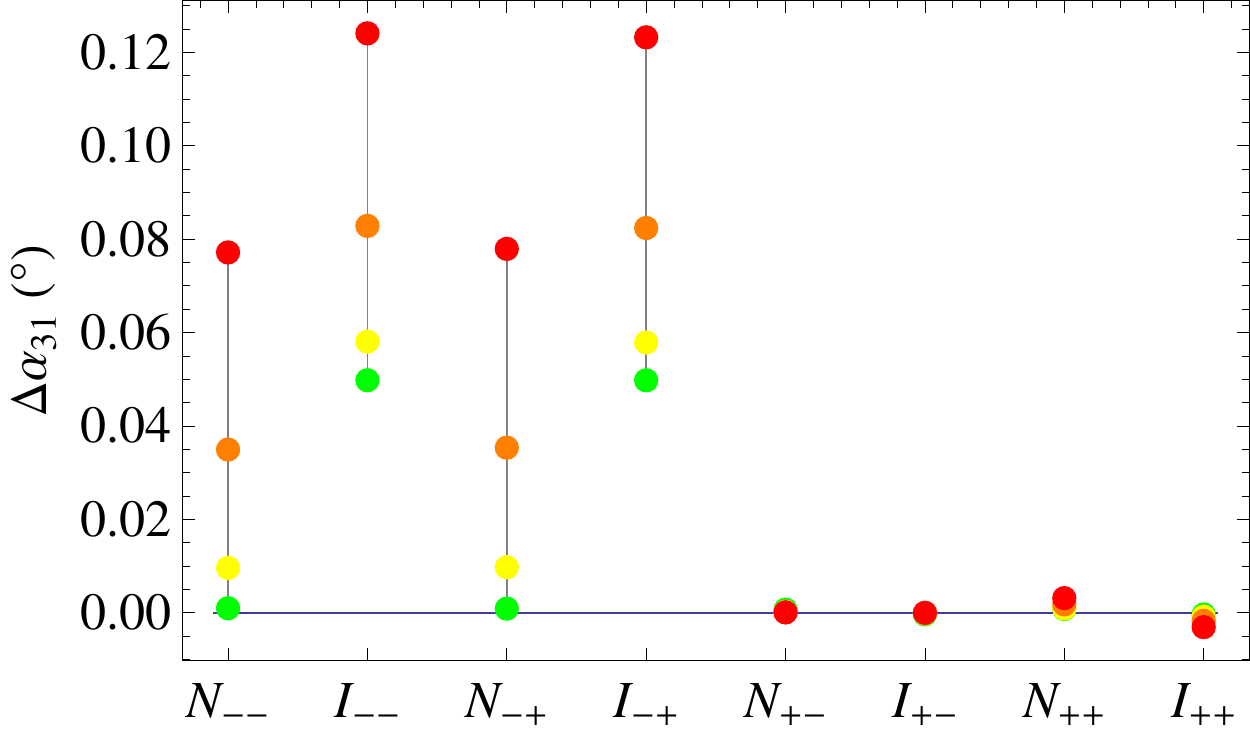}
\end{minipage}

\caption{\label{fig:8cases}RG corrections for all the 8 cases, $N$/$I$ for
normal/inverted hierarchy and ``$+/-$'' for $e^{i\alpha}=+1/-1$
where $\alpha$ stands for the two Majorana phases. The result depends
on the lightest neutrino mass, which is set at $(1,\thinspace20,\thinspace40,\thinspace60)$
meV for points colored from green to red.}
\end{figure}

Therefore, there are four positive/negative signs (and thus $16$
physically inequivalent cases) relevant in determining $(r,\thinspace x_{1},\thinspace x_{2},\thinspace y)$.
However, as it can be seen from the mass matrix, for $\delta=+90^{\circ}$
and $-90^{\circ}$, the mass matrix in one case is simply the complex
conjugate of the other, so we only need to study one of the two cases.
Actually,  the result of RG running of both cases shows that the
radiative corrections on both cases are the same except that for $\delta$
it differs by a minus sign. This reduces the 16 cases to 8 cases in
our analysis. In addition, the case of positive $\delta=+90^{\circ}$
is disfavored by current global fits.
For simplicity, we refer to the 8 cases
as $N_{\pm\pm}$ and $I_{\pm\pm}$ where $N$/$I$ stands for the
normal/inverted ordering and the two $\pm$ stand for the signs
of $e^{i\alpha_{21}}$ and $e^{i\alpha_{31}}$, respectively.

We solve the RGEs using the code REAP \cite{Antusch:2005gp} and compute
the RG corrections.  The results are presented in Fig.\ \ref{fig:running}
for $N_{++}$ and Fig.\ \ref{fig:8cases} for all the 8 cases. We set
the flavor symmetry scale at $\Lambda=10^{14}$ GeV. Actually as shown
in Fig.\ \ref{fig:running} the RG corrections depend linearly on $\log\Lambda$,
so if $\Lambda$ is changed to another value $\Lambda'$, the RG corrections
can be evaluated correspondingly by simply multiplying a factor of
$\log\Lambda'/\log\Lambda$. Another parameter that may have significant
effect  is the lightest neutrino mass $m_{L}$. In Fig.\ \ref{fig:8cases}
we show the RG corrections for different values of $m_{L}$ by green,
yellow, orange and red points, corresponding to $m_{L} = (1,\thinspace20,\thinspace40,\thinspace60)$
meV respectively\footnote{We do not take $m_{L}=0$ here because for $m_{L}=0$, the RG corrections
are almost the same as $m_{L}=1$ meV except for the Majorana phases
which are not well defined when $m_{L}=0$. }. We assume here that
strong limits on the neutrino mass scale from cosmology are valid
\cite{Archidiacono:2016lnv} and simply note that the effect of running roughly scales with
$m_L$ for values larger than 60 meV.

As shown in Fig.\ \ref{fig:8cases}, typically the corrections to $\theta_{12}$,
$\theta_{13}$, $\theta_{23}$, $\delta$, $\alpha_{21}$, $\alpha_{31}$
are about $0.1$, $0.001$, $0.005$, $0.1$, $0.05$, $0.1$ degrees
respectively, except for some cases where due to some cancellations
the RG corrections are suppressed. To understand the cancellation, we take
$\theta_{12}$ as example, for which the analytic expression reads \cite{Antusch:2003kp}
\begin{equation}
\frac{d\theta_{12}}{d\ln\mu}=-\frac{y_{\tau}^{2}}{32\pi^{2}}\sin2\theta_{12}s_{23}^{2}\frac{|m_{1}+m_{2}e^{i\alpha_{21}}|^{2}}{\delta m^{2}}+{\cal O}(\theta_{13}).\label{eq:mt-40}
\end{equation}
Here $y_\tau$ is the tau-lepton Yukawa coupling.
The plot for $\theta_{12}$ in Fig.\ \ref{fig:8cases} shows that
the corrections in the four cases $N_{-\pm}$ and $I_{-\pm}$ are
suppressed, which can be understood from Eq.~(\ref{eq:mt-40}):
the correction is proportional to $|m_{1}+m_{2}e^{i\alpha_{21}}|^{2}$, which
can be small if $e^{i\alpha_{21}} = -1$ and $m_1 \approx m_2$. The latter always happens in the
inverted ordering and in the normal ordering when the smallest mass $m_1$ approaches
$\sqrt{\delta m^{2}}$.

Except for some  cases with cancellations, the RG corrections generally
increases when $m_{L}$ increases. This behavior is very common regarding
small perturbations to the mass matrix, which has been studied in
Ref.\ \cite{Rodejohann:2015nva} from a more general point of view.
The reason is because for larger
$m_{L}$, the mass spectrum is closer to the quasi-degenerate situation,
where the PMNS mixing becomes unstable when the mass matrix suffers
perturbations. Besides, among the three mixing angles, $\theta_{12}$
generally receives the largest correction (except for cancellations);
this is because the gap between $m_{1}$ and $m_{2}$ is much smaller
than that of $m_{1}$ and $m_{3}$ or $m_{2}$ and $m_{3}$.

\begin{figure}
\centering

\begin{minipage}{0.42\textwidth}
\flushright
\includegraphics[height=4cm]{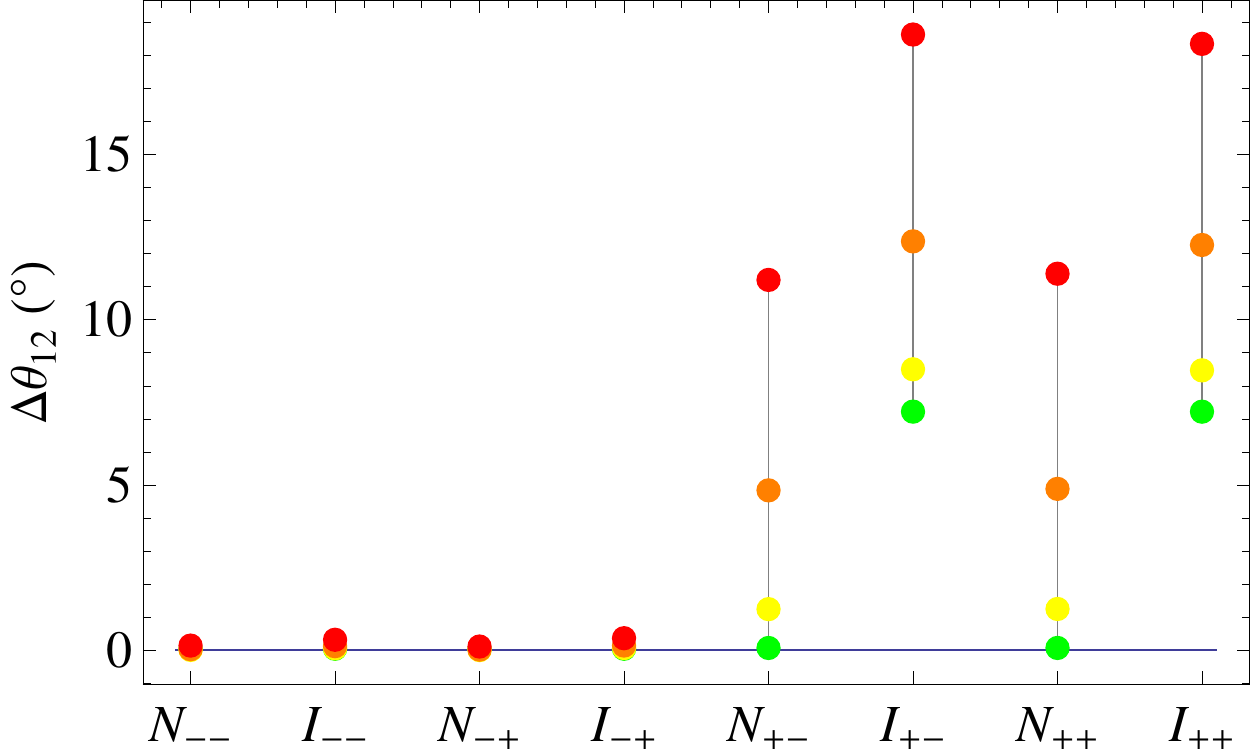}\\
\includegraphics[height=4cm]{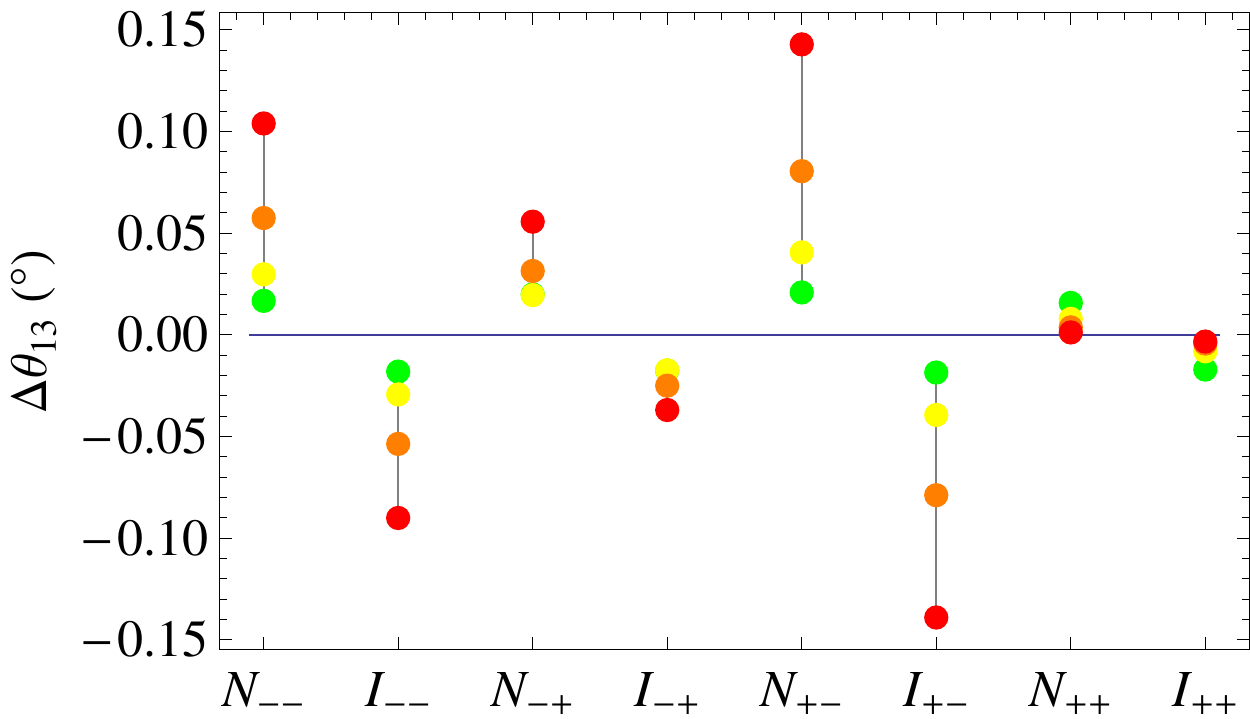}\\
\includegraphics[height=4cm]{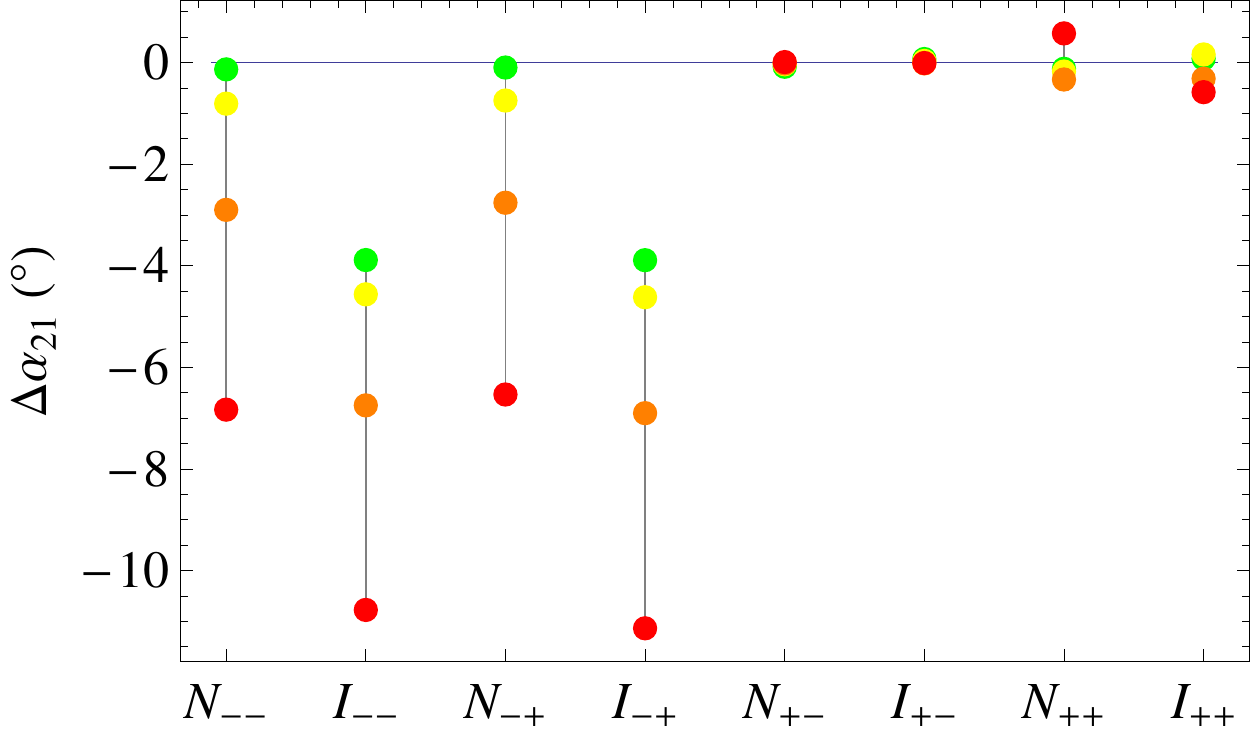}
\end{minipage}
\begin{minipage}{0.42\textwidth}
\flushright
\includegraphics[height=4cm]{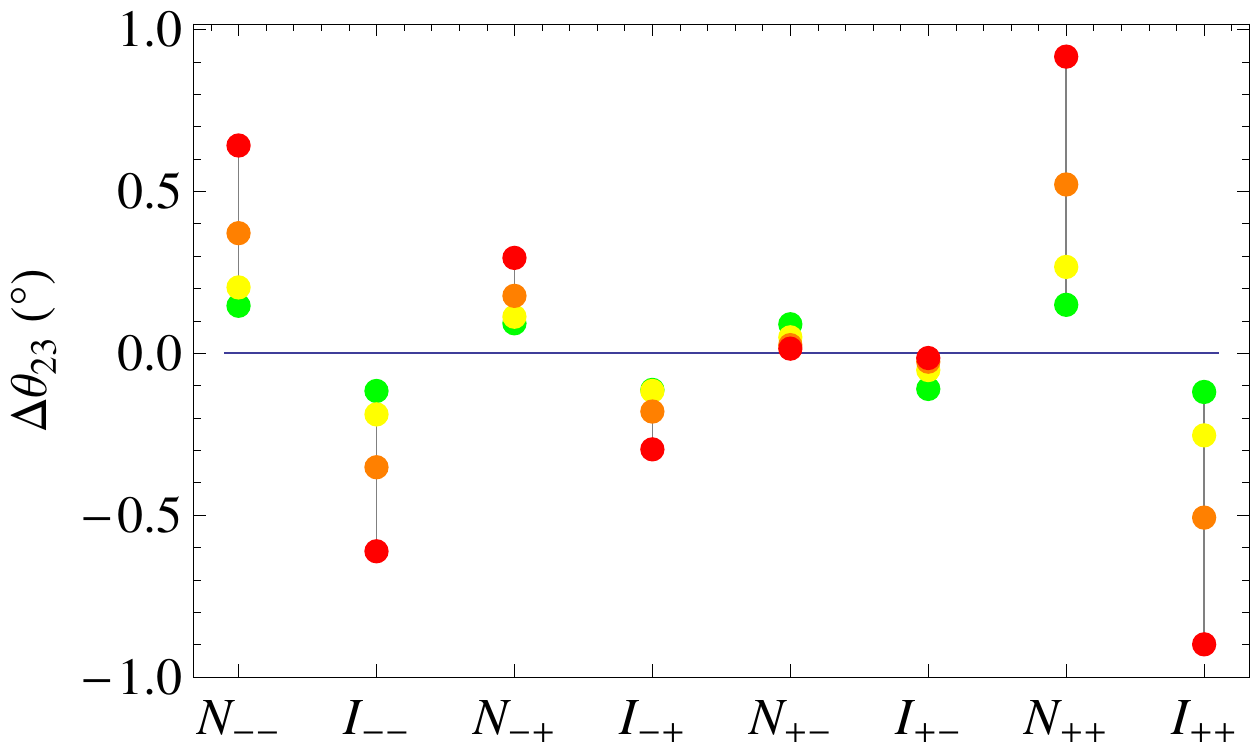}\\
\includegraphics[height=4cm]{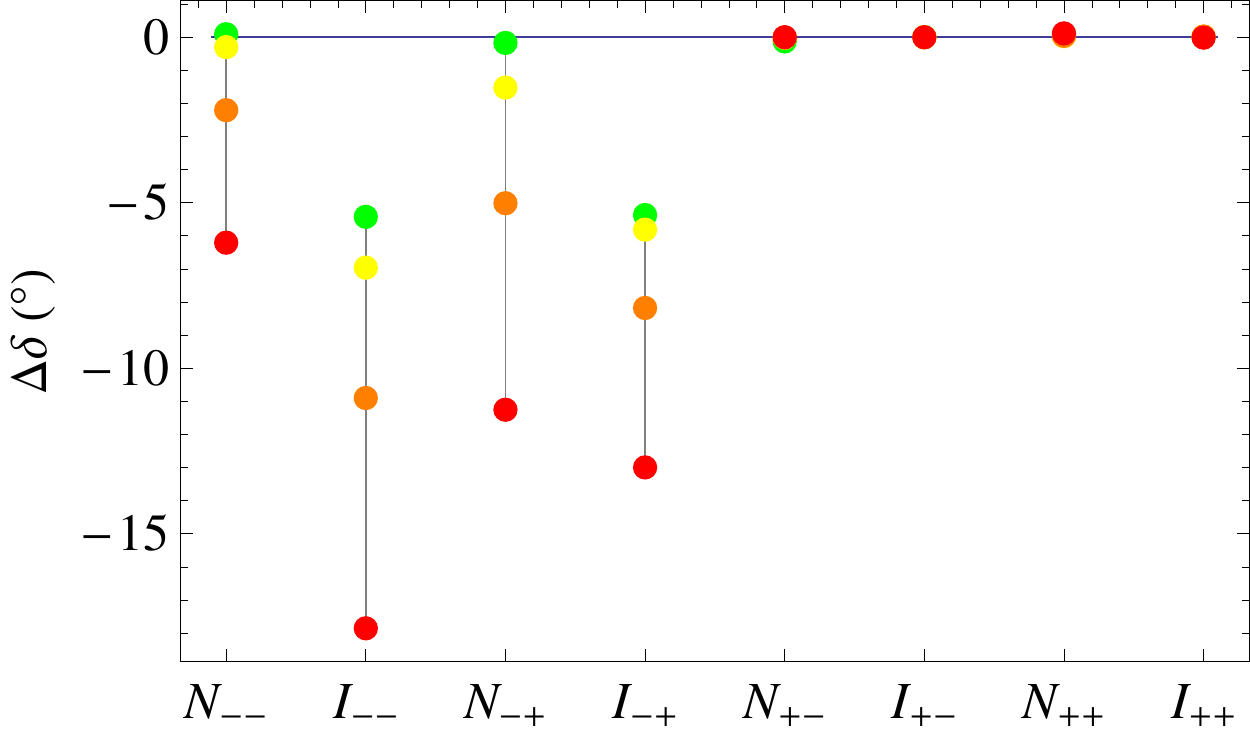}\\
\includegraphics[height=4cm]{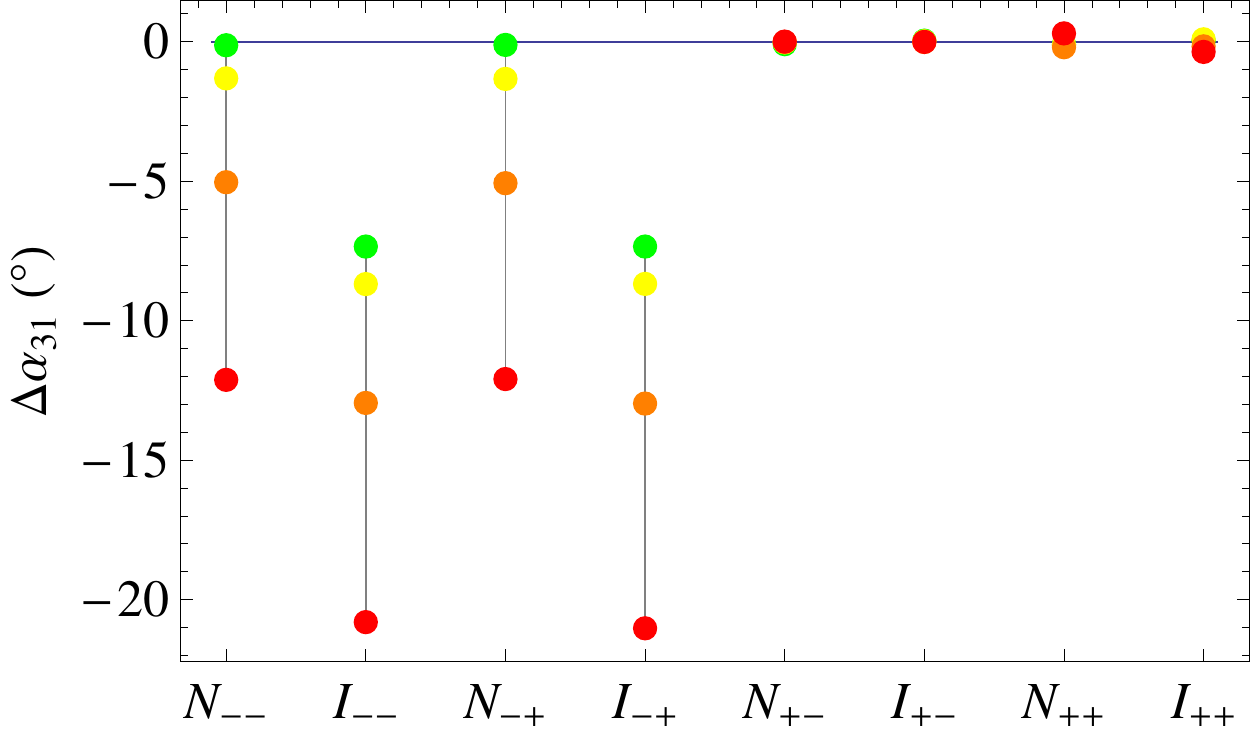}
\end{minipage}

\caption{\label{fig:8casesMSSM}Same as Fig.\ \ref{fig:8cases} except that the SM is replaced with
the MSSM for $\tan \beta = 20$.}
\end{figure}

Since all the corrections are at the order of or even lower than $0.1^{\circ}$,
we can draw the usual conclusion that in the context of the SM with the
Weinberg operator only, the RG corrections are negligible when  compared
with current and near future experimental measurements. As well known, if we replace the SM with the
MSSM, then according to  Ref.\ \cite{Antusch:2003kp} the RG corrections to the neutrino mixing
would be amplified by a factor of $\tan^{2}\beta$. To illustrate
this effect, we compute the RG corrections again in the context of the MSSM with
$\tan \beta=20$, and the result is shown in Fig.\ \ref{fig:8casesMSSM}.
As one can see, the RG corrections in the MSSM with large $\tan\beta$ are significantly
enhanced to measurable values compared to Fig.\ \ref{fig:8cases}.

\subsection{RG running based on type I seesaw \label{sec:RG-running-typeI}}

In this section, we consider new physics that appears below the flavor
symmetry scale. The Weinberg operator itself is UV incomplete and
is usually believed to be a low-energy effective operator.
Here we consider the type I seesaw realization of this operator only. Heavy
right-handed neutrinos $N_{i}$ ($i=1,\thinspace2,\ldots$)
are integrated out to generate the Weinberg operator.
We consider the scenario that the right-handed neutrino masses (or
the seesaw scale) are lower than the flavor symmetry scale. So at
the flavor symmetry scale, we should consider the symmetry of the
following Lagrangian instead of the Weinberg operator,
\begin{equation}
{\cal L}\supset-y_{ij}N_{i}\widetilde{H}^{\dagger}L_{j}-\frac{1}{2}N_{i}M_{ij}N_{j}+\rm{h.c}.\label{eq:mt-36}
\end{equation}

Next we need to specify the transformation rules of $\mathbb{Z}_{2}^{{\rm TM1}}$
and $\mathbb{Z}_{2}^{\rm CP}$ for the right-handed neutrinos. This depends
on how we assign the right-handed neutrinos to the representations
of the flavor symmetry, which is rather model-dependent.
For simplicity, we assume that the number of right-handed neutrinos is
three and that they have the same transformation rule as the left-handed
neutrinos. As a result, both the Dirac mass matrix $m_{D}$ and the heavy Majorana matrix $M$ will be in
the form of Eq.~(\ref{eq:mt-13}). As one can check explicitly, if
both $m_{D}$ and $M$ are in the form of Eq.~(\ref{eq:mt-13}), then
the light-neutrino mass matrix
\begin{equation}
M^{\nu}=-m_{D}^{T}\,M^{-1}\,m_{D}\label{eq:mt-37}
\end{equation}
is also of the form in Eq.~(\ref{eq:mt-13}). As we have discussed,
each matrix of the form (\ref{eq:mt-13}) contains four real parameters
thus in the Lagrangian (\ref{eq:mt-36}) we have 8 free parameters.
 The tree-level predictions in Eqs.~(\ref{eq:mt-28}) and (\ref{eq:mt-29})
are independent of the values of these parameters. However, the RG corrections
inevitably depend on these parameters. As we have argued, when some
new physics such as the right-handed neutrinos appears below the flavor
symmetry scale, the RG corrections would usually depend on many unknown
parameters, which makes it difficult to evaluate the RG corrections
exactly. To understand generally how large the RG corrections would
be, we adopt random scattering in the allowed parameter space rather
than focus on some specific parameter settings.

\begin{figure}
\centering

\includegraphics[width=5cm]{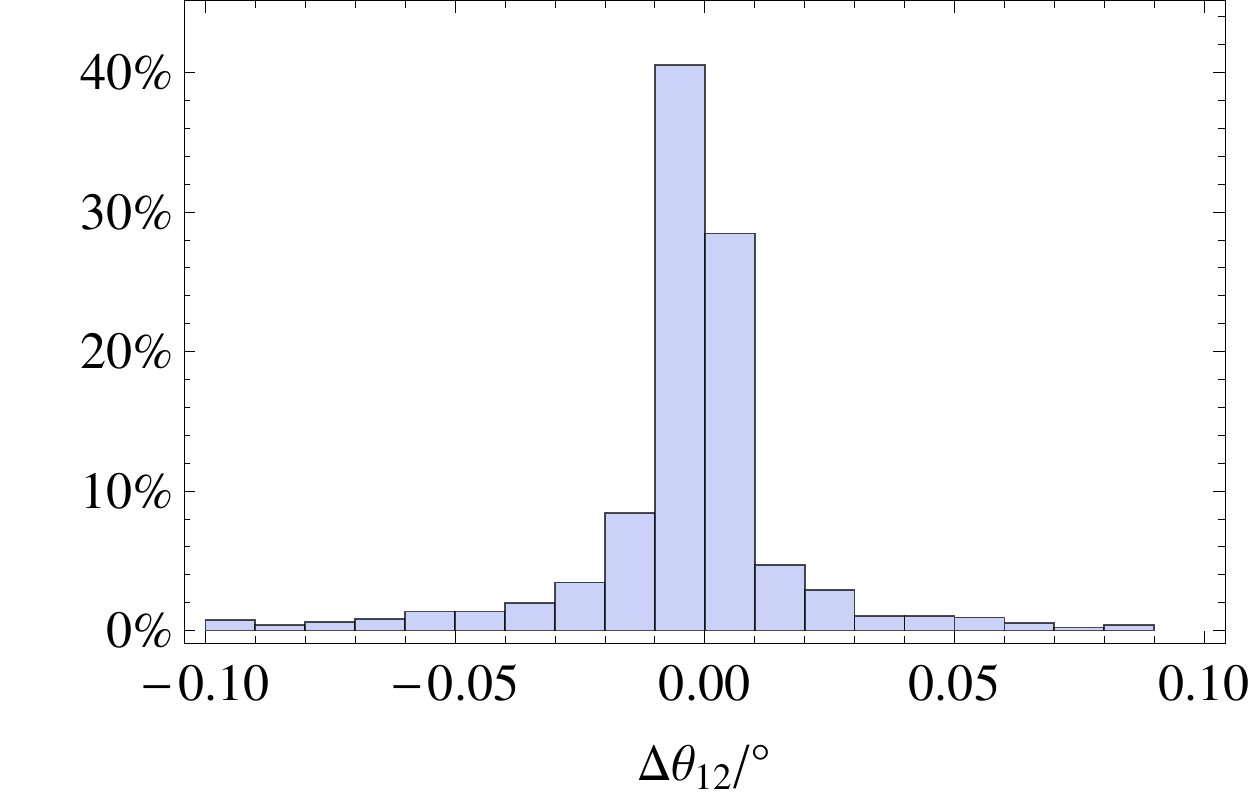}\,\includegraphics[width=5cm]{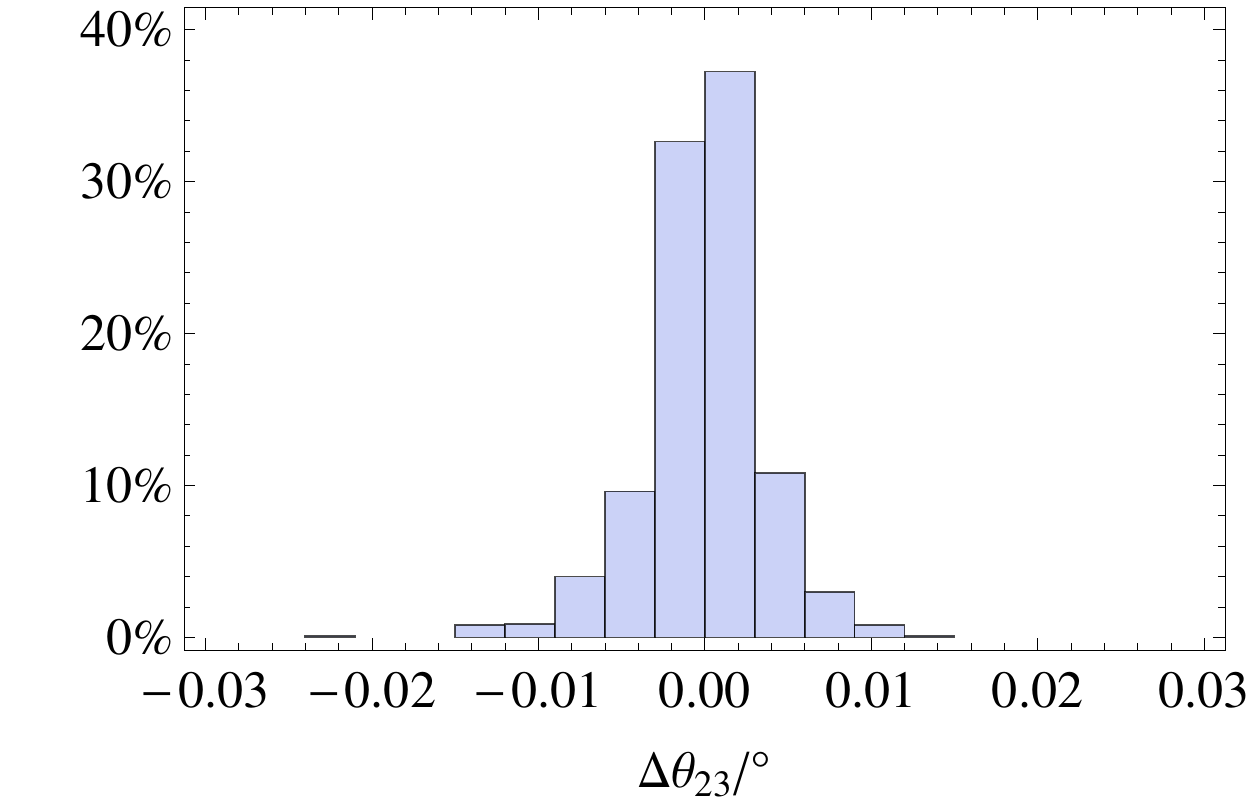}\,\includegraphics[width=5cm]{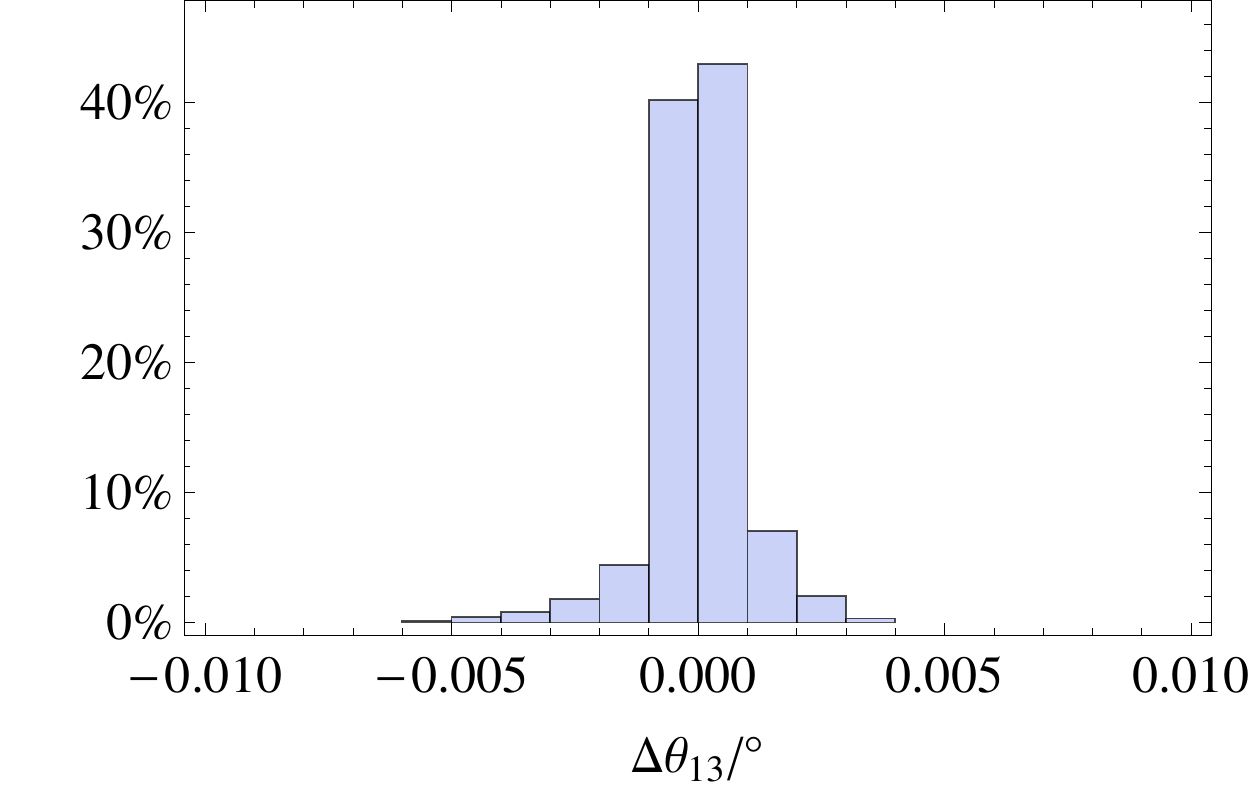}

\vspace{0.5cm}

\includegraphics[width=5cm]{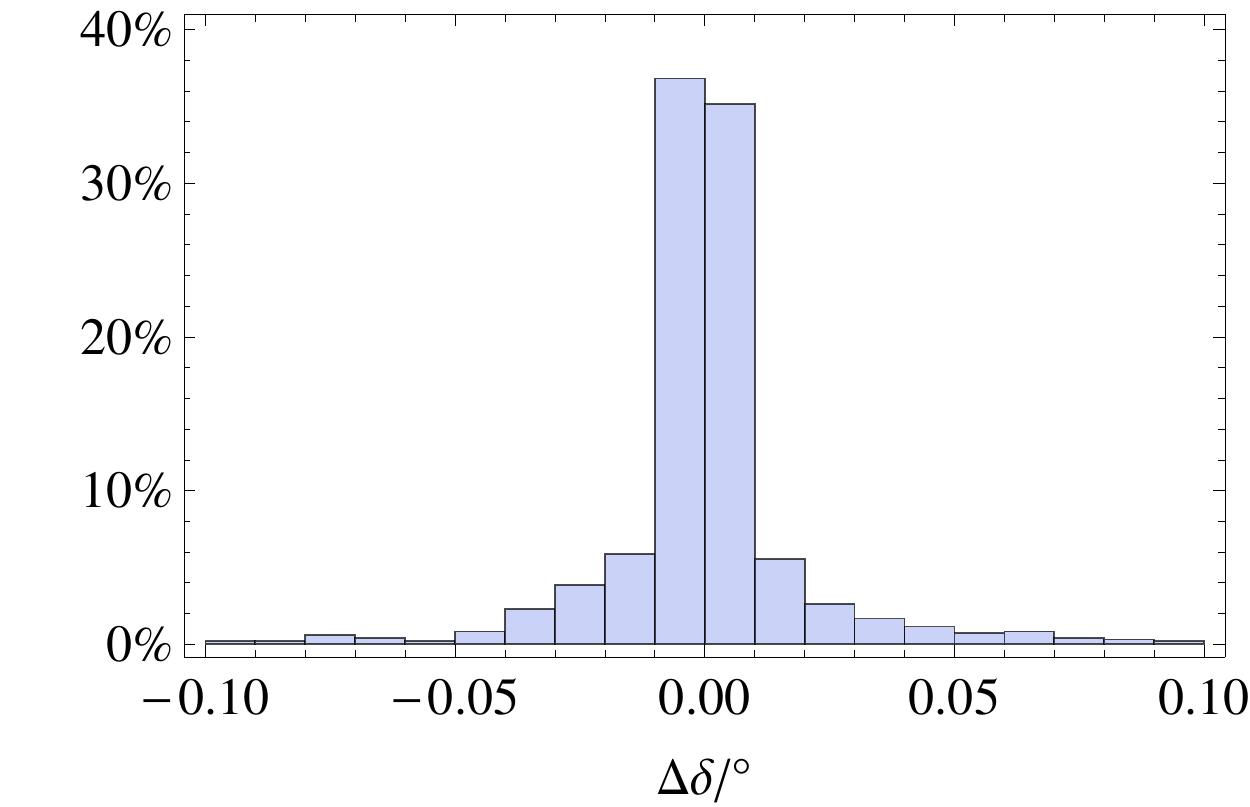}\,\includegraphics[width=5cm]{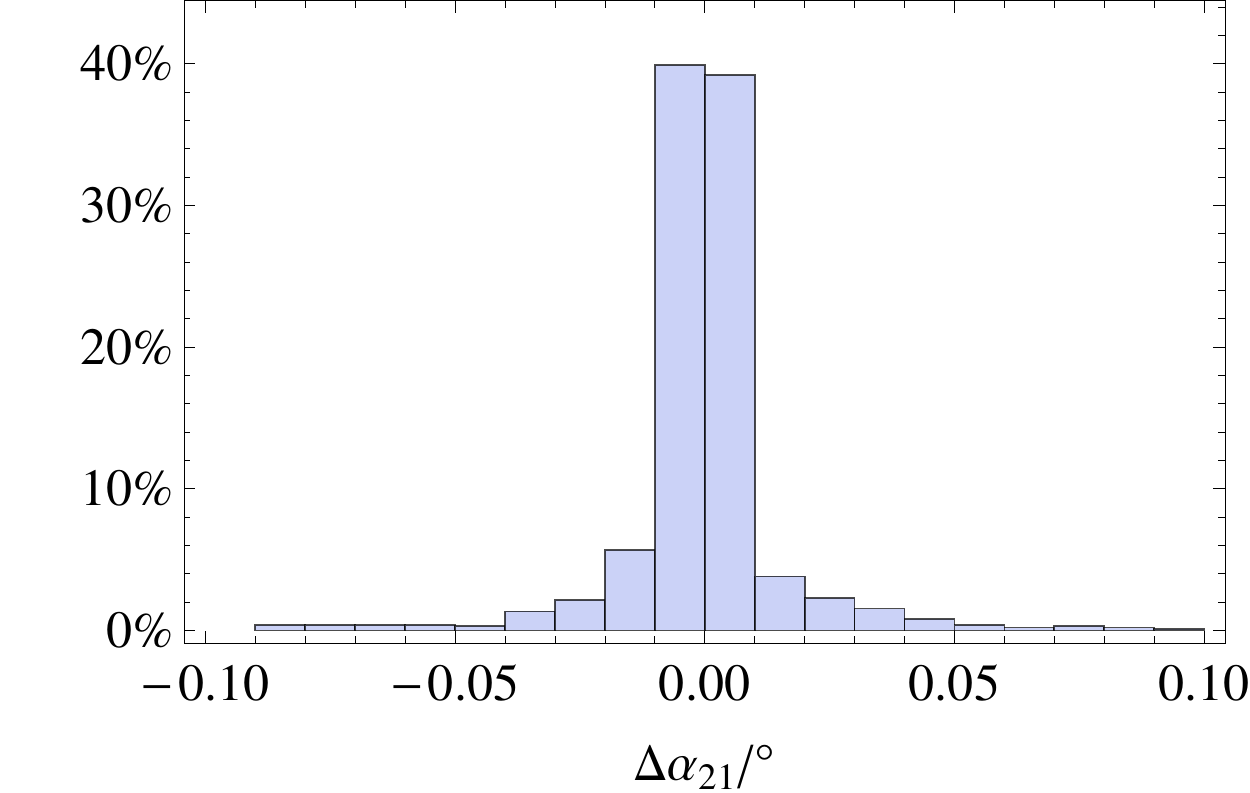}\,\includegraphics[width=5cm]{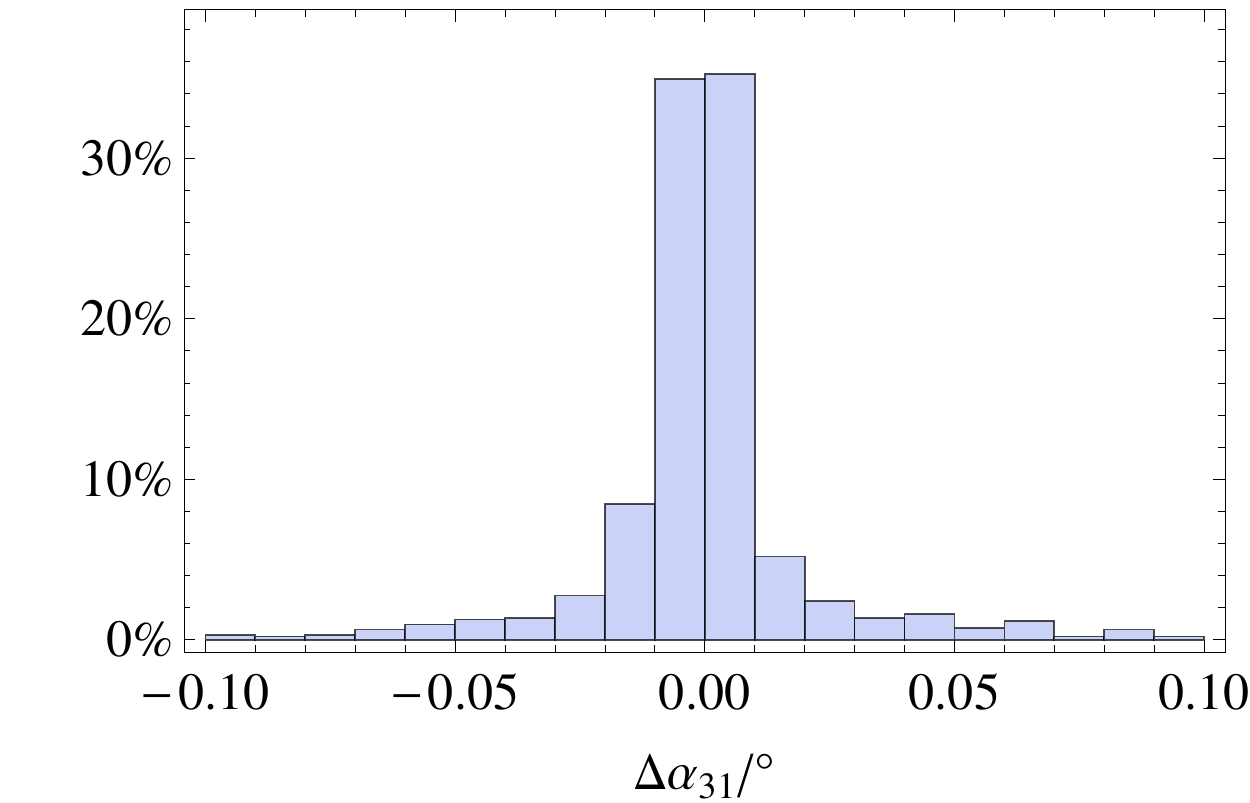}\caption{\label{fig:distribution}The distributions of RG correction in the
SM extended by the type I seesaw. }

\end{figure}

We randomly generate 1000 samples with right-handed neutrino masses
$M_{1}$, $M_{2}$, $M_{3}$ distributed from $10^{6}$ GeV to $10^{13}$
GeV and the lightest neutrino mass $m_{L}$ from 1 meV to 60 meV. Rectangular
distributions are used for $\log M_{1,2,3}$ and $m_{L}$. The positive/negative
signs of $\Delta m^{2}$ and Dirac/Majorana phases are also chosen
randomly. The Yukawa couplings can be computed once $(M_{1},\thinspace M_{2},\thinspace M_{3})$
and $(m_{1},\thinspace m_{2},\thinspace m_{3})$ have been set. We
again use the code REAP \cite{Antusch:2005gp}, which automatically integrates
out the heavy right-handed neutrinos when the energy scale goes below
their masses.

The results are presented in Fig.\ \ref{fig:distribution}, where
we can see most RG corrections are distributed in small ranges, e.g.
$\Delta\theta_{12}$, $\Delta\theta_{23}$ and $\Delta\theta_{13}$
are most likely less than $0.05^{\circ}$, $0.01^{\circ}$ and $0.005^{\circ}$
respectively. So generally, the deviations are similar to the results
in Fig.\ \ref{fig:8cases} where right-handed neutrinos are not introduced.
However, large corrections are also possible. We do not find any significant
cut-off of the deviations when the number of samples are increased,
though the distributions above remain almost the same. This implies
the RG corrections could be very large, but would require
fine-tuning in the parameter space. For example, when the number of samples is increased to $10^{4}$,
we find only two samples with $|\Delta\theta_{23}|>3{}^{\circ}$.
Therefore,
we can draw the conclusion that generally the RG corrections
in the type I seesaw scenario are of similar magnitude as  with the Weinberg
operator only.

\begin{figure}
\centering

\includegraphics[width=5cm]{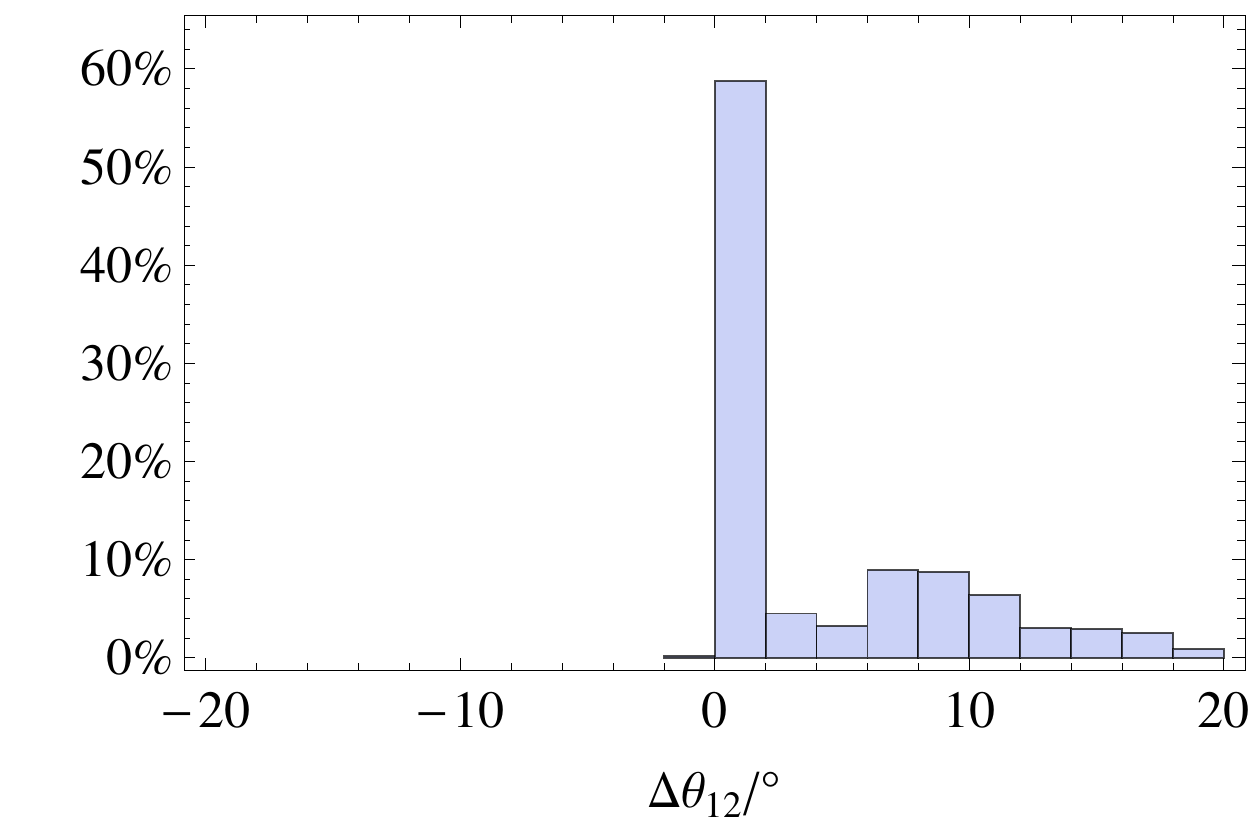}\,\includegraphics[width=5cm]{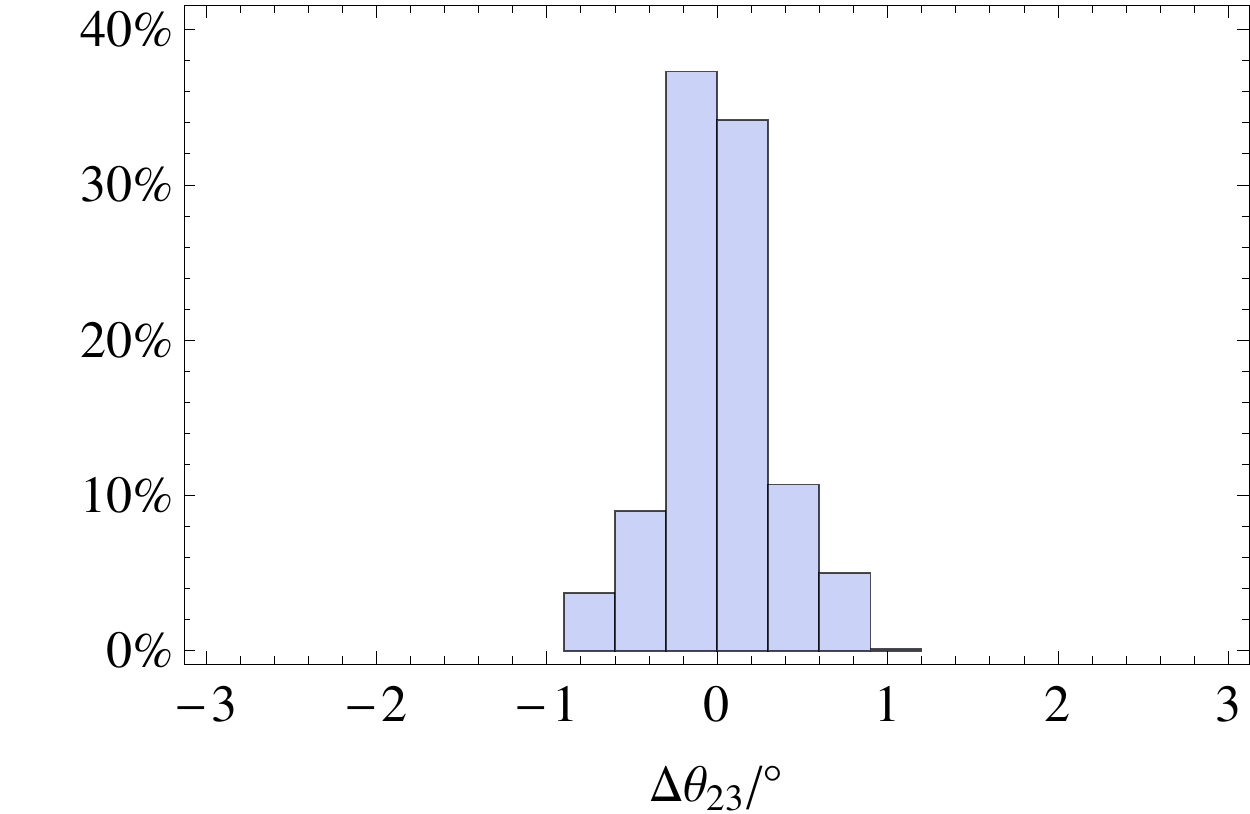}\,\includegraphics[width=5cm]{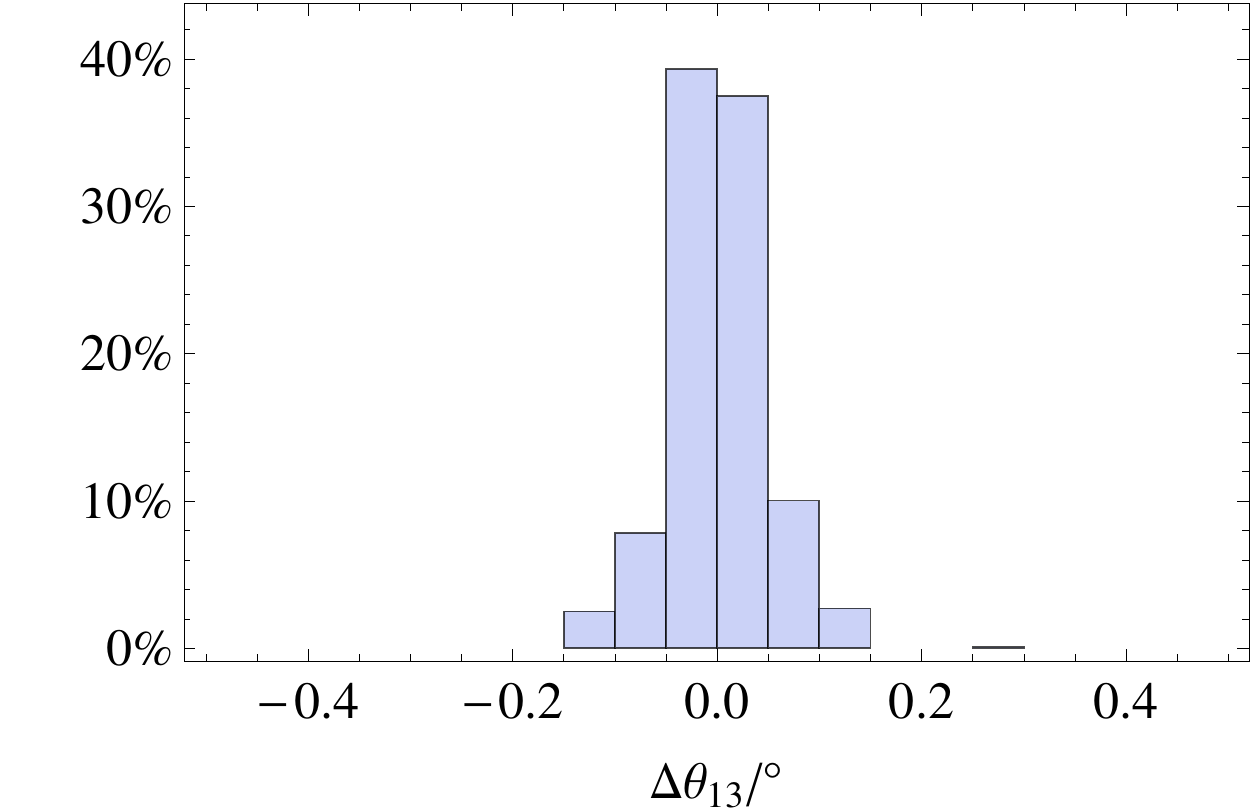}

\vspace{0.5cm}

\includegraphics[width=5cm]{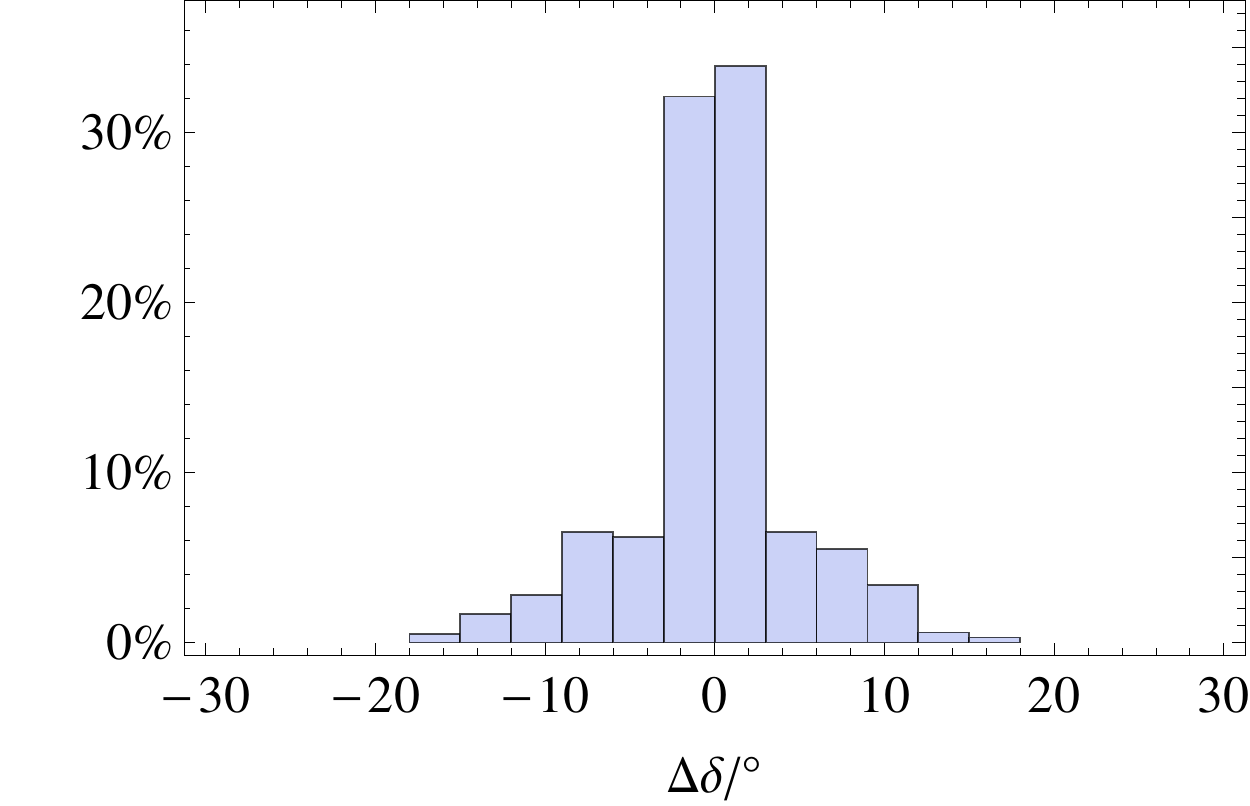}\,\includegraphics[width=5cm]{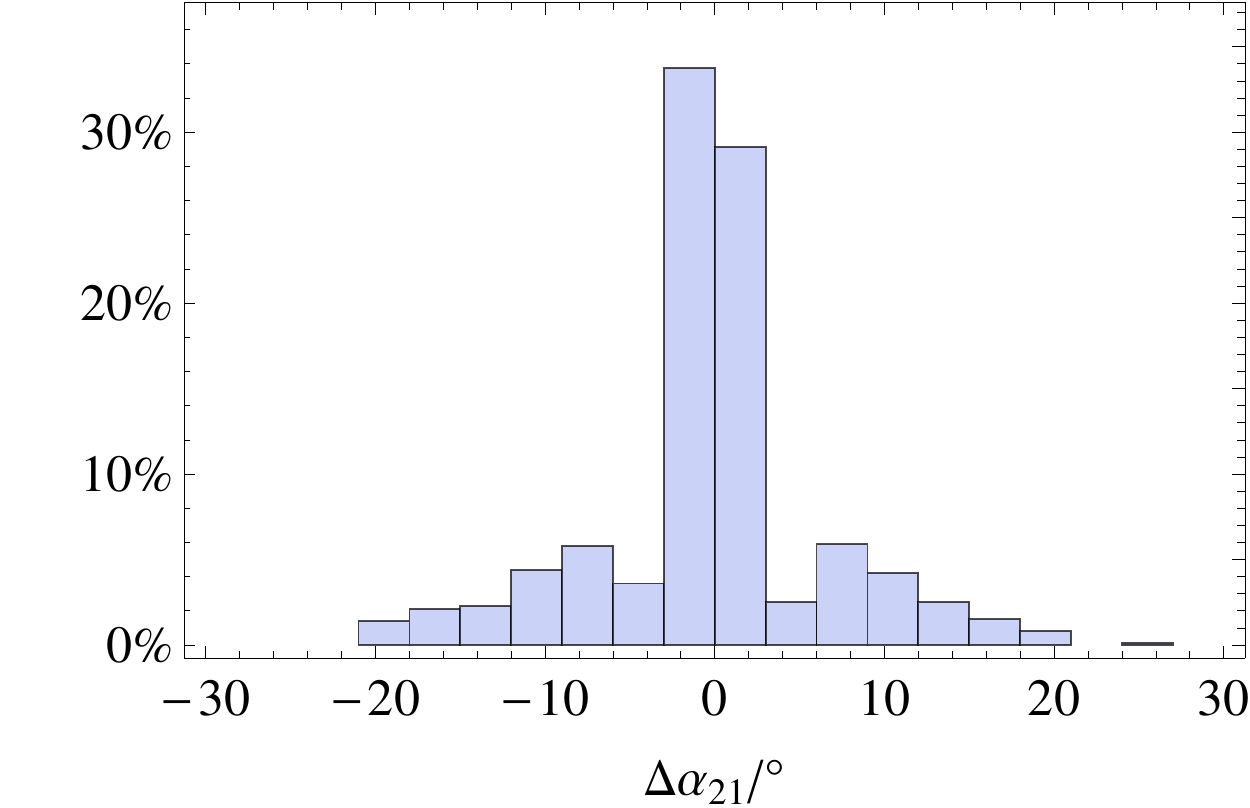}\,\includegraphics[width=5cm]{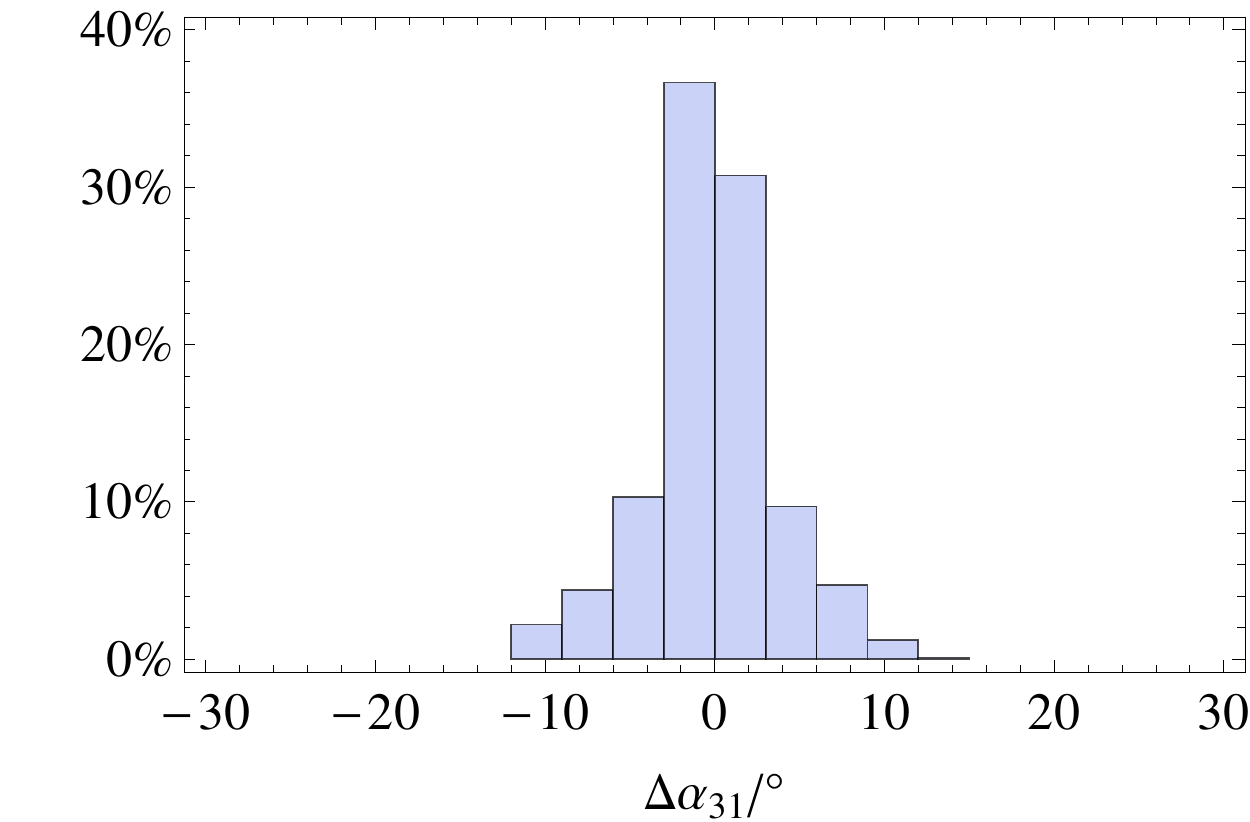}\caption{\label{fig:distribution-mssm}The distributions of RG correction
in the MSSM with $\tan \beta = 20$ extended by the type I seesaw.
}
\end{figure}

Again, the RG corrections
can be significantly amplified within supersymmetric scenarios.
We compute the RG corrections in the MSSM extended by the type I seesaw
with $\tan\beta=20$. The result is shown in Fig.\ \ref{fig:distribution-mssm}
where we can see that compared to Fig.\ \ref{fig:distribution},
the RG corrections in the MSSM with large $\tan\beta$ are significantly
enhanced by up to two orders of magnitude to measurable values.

\section{Conclusion\label{sec:Conclusion}}

Combining $\mu$-$\tau$ reflection and TM1 symmetry leads to a very predictive framework. We
have shown in Sec.\ \ref{sec:basic} that it not only can accommodate
non-zero $\theta_{13}$ but also predicts all other PMNS parameters, including all CP phases
($\delta = \pm \pi/2$ and the Majorana phases are $0$ or $\pi$).

With these symmetries, the neutrino mass matrix can be constrained
to the form (\ref{eq:mt-13}) containing only four real parameters.
Given the experimental values of $\theta_{13}$, $\delta m^{2}$ and $\Delta m^{2}$
as input, the mass matrix can be exactly reconstructed for a fixed
value of the smallest mass $m_{L}$ and several choices of positive/negative signs.
Therefore, for the SM extended by the Weinberg operator, the RG corrections
can be exactly evaluated as the only free parameter is $m_{L}$.

We have computed the RG corrections to the scenario, which are in agreement with
known results, namely that in the SM they are typically small, but can
be enhanced to measurable values
within supersymmetric scenarios and within explicit multi-scale scenarios such as the
type I seesaw mechanism.

In summary, the mixing scheme we propose here is very well compatible with data and
addresses the closeness of $\delta$ with $-\pi/2$, of $\theta_{23}$ with $\pi/4$ and that
$\sin^2 \theta_{12}$ is slightly less than $1/3$. If future data confirms those special values of
the mixing parameters, the proposed scheme seems an attractive approach to the description of lepton mixing.
On the other hand, some deviations could occur in the future, which
could either be explained by RG corrections if the deviations are
small, or exclude this mixing scheme if they are large. One particularly
noteworthy example is the deviation of $\theta_{23}$ from $45^{\circ}$,
which was recently hinted by the NOVA measurement \cite{Adamson:2017qqn} $\theta_{23}=39.5^{\circ}{}_{-1.3}^{+1.7}$
or $52.2^{\circ}{}_{-1.8}^{+1.3}$ . Such a large deviation $(\gtrsim5^{\circ})$
if confirmed by future data, would exclude this mixing scheme embedded
in the simple scenarios considered in this paper.



\begin{acknowledgments}
We thank Stefan Bruenner and Ludwig Rauch for helpful discussions.
WR is supported by the DFG with grant RO 2516/6-1 in
the Heisenberg program.
\end{acknowledgments}

\bibliographystyle{apsrev4-1}
\bibliography{ref}

\end{document}